# SHACL validation in the presence of ontologies: Semantics and rewriting techniques

Anouk Oudshoorn *, Magdalena Ortiz, Mantas Šimkus

*TU Wien, Austria*

A B S T R A C T

SHACL and OWL are two prominent W3C standards for managing RDF data. These languages share many features, but they have one fundamental difference: OWL, designed for inferring facts from incomplete data, makes the *open-world assumption*, whereas SHACL is a constraint language that treats the data as complete and must be validated under the *closed-world assumption*. The combination of both formalisms is very appealing and has been called for, but their semantic gap is a major challenge, semantically and computationally. In this paper, we advocate a semantics for SHACL validation in the presence of ontologies based on *core universal models*. We provide a technique for constructing these models for ontologies in the rich data-tractable description logic Horn-$\mathcal{ALCHIQ}$. Furthermore, we use a finite representation of this model to develop a *rewriting technique* that reduces SHACL validation in the presence of ontologies to standard validation. Finally, we study the complexity of SHACL validation in the presence of ontologies, and show that even very simple ontologies make the problem ExpTime-complete, and PTime-complete in data complexity.

## 1. Introduction

The Shape Constraint Language (SHACL) [1] and Web Ontology Language (OWL) [2] are two prominent W3C standards for managing RDF data, a graph-based data model of the Web [3]. These standards are based on fundamentally different assumptions and designed to be complementary. OWL was standardised shortly after RDF, with the key aim of enhancing RDF datasets with domain knowledge that enables the inference of missing facts from potentially incomplete data graphs. OWL and its profiles are based on *Description Logics (DLs)* [4] and, like other classical logics, make the *open-world assumption (OWA)*, which intuitively means that the data only presents an incomplete description of the domain of interest: it asserts facts that are known to be true, but does not rule out that additional facts may also be true, as long as they are consistent with the current world description. OWL has been adopted in a wide range of applications over the years, and thousands of OWL ontologies have been developed. SHACL, in contrast, was created for a different purpose: to describe and validate constraints on datasets. The main task of interest is *validation* of a given a set of constraints paired with a selection of *target* nodes or concepts from a given graph. Unlike OWL, SHACL operates under the *closed-world assumption* (CWA): it assumes that the given data graph is complete, and validators evaluate the constraints over the input graph as is.

A natural question is how to do validation in the presence of both OWL ontologies and SHACL constraints. That is, if we have a possibly incomplete graph and ontological knowledge that implies additional facts, can we validate given SHACL constraints over graphs containing the implied facts? Consider as an example a toy database of pet owners containing the facts $hasPetBird(linda, blu)$, $hasPet(john, ace)$; the simple constraint $petOwnerShape \leftarrow \exists hasPet.\top$, which says that everyone that has a pet is a pet owner; and the target pet owners $linda$ and $john$. Clearly, we would like to leverage the knowledge that *all pet birds are pets*, written as $hasPetBird \sqsubseteq hasPet$ in description logics, which allows us to validate both targets. This type of validation is very natural, even more so in the light of the large amount of ontologies that are already being used for describing data on the web. It is in fact envisioned in the W3C SHACL specification, which calls for graph validation in the presence of OWL entailment [1, Section 1.5], but unfortunately, does not provide guidance on how to realise this. Nevertheless, there are approaches of reconciling the two [5,6], and concrete ontology and constraint pairs being constructed [7].

* Corresponding author.
*E-mail addresses:* anouk.oudshoorn@tuwien.ac.at (A. Oudshoorn), magdalena.ortiz@tuwien.ac.at (M. Ortiz), mantas.simkus@tuwien.ac.at (M. Šimkus).

The first major challenge we must face is that the semantics of SHACL constraints in the presence of ontologies is not obvious, as we must simultaneously account for the open-world semantics of description logics and for the closed-world view of SHACL. The knowledge in the ontology implies additional facts, which can be added to the data graph so that it satisfies all the ontological axioms. All such possible completions are *models* that must be taken into account according to the traditional OWL/description logics semantics. However, SHACL allows for negation, which makes this *certain answer semantics* too weak, and quickly results in non validation. Update for example the toy database of pet owners we considered before to $hasPet(john, ace)$, $Hamster(ace)$, but with the constraint $petOwnerShape \leftarrow \exists hasPet.\top \wedge \forall hasPet.\neg Dangerous$. Like before, this expresses that pet owners own a pet. Moreover it is required that all pets are not dangerous. For $john$ to validate the $petOwnerShape$, it is not enough to make explicit in the ontology that hamsters are not dangerous animals; it should also be enforced that any possible pet $john$ might have cannot be dangerous.

For lightweight DLs, fragments of classical Horn logics that cannot express disjunctive information, *universal models* can be obtained using standard, database-inspired *chase procedures*. These models can be used for evaluating conjunctive, navigational and graph queries in the presence of ontologies, see [8–13] and their references. One of these options, using *minimal models* of the Skolemnisation of the DL ontology, has been advocated for in the case of integrity constraints [6,14]. But even such models give very weak semantics in formalisms with negation such as SHACL. Let us for example consider again another version of the toy database containing pet owners: assume it contains the facts $hasPet(john, ace)$, $PetOwner(john)$, $Hamster(ace)$. It is conceivable to find an axiom like $PetOwner \sqsubseteq \exists hasPet.\top$ in the accompanying ontology. Now let the $onlyHamsterShape$ be given by the constraint enforcing all pets to be hamsters, $onlyHamsterShape \leftarrow \forall hasPet.Hamster$, which we want to validate for $john$. This is clearly the case for the original database. It is also true that the given database is already satisfying all given ontology axioms, so it seems there is no reason to change the validation result. However, the minimal model under the Skolem semantics adds a fresh node as a $hasPet$-child of $john$, without the label $Hamster$, changing the validation result to negative.

To obtain stronger and more intuitive semantics, and to avoid the problems presented in the previous example, we advocated in [15] for an *austere* canonical model in which axioms are satisfied minimally, introducing as few successors as possible without losing universality. We showed that for ontologies in *DL-Lite*$_{\mathcal{R}}$—the logic underlying OWL 2 QL [16]—such a model can be represented by a so-called *immediate successor function* that describes the minimal set of facts that need to be added to satisfy the axioms at a given point of the model construction. The model itself can then be obtained in a deterministic, step-by-step fashion. We extend this construction in Section 3 to the significantly more expressive Horn-$\mathcal{ALCHIQ}$, one of the largest fragments of OWL that is still contained in Horn logic. Crucially, we show that the resulting model is a *core* in the traditional database sense. This provides strong evidence in favour of our chosen semantics, since cores are often advocated as the adequate choice for languages that are not closed under homomorphisms, but satisfy the weaker property of being closed under isomorphisms [17–19]. In the light of this relationship, our austere model construction provides a novel technique for building core models without the expensive core-checking step of traditional core chase procedures. As we point out, the same applies to some previous model constructions from the DL literature [11,18].

With our semantics based on austere (i.e., core) models in place, we can tackle the problem of devising an algorithm for validation. Constructing the austere model may be infeasible in practice, since it is infinite in general. Instead, we use our finite representation of the model. Ideally, we would like to realise validation via *rewriting*. That is, we want to compile a given ontology and a set of SHACL constraints into a new set of SHACL constraints that incorporate the relevant knowledge of the ontology in such a way that the implicit facts are taken into account in validation, without having to explicitly add them to the graph. Rewriting techniques are very desirable as they open the way to reuse standard SHACL validators to perform validation in the presence of ontologies. We use the finite representation of the austere canonical model to construct a complex structure that stores so-called *2-types*, which intuitively represent an abstract copy of a possible object and its neighbours in the austere model of a given data graph—enriched with information about the shapes that are (not) satisfied in implied substructures. This structure is used to induce a modified set of SHACL constraints that validate over a given data graph exactly when the original constraints validate over the austere canonical model of the input graph and the ontology. We first develop the technique for a positive fragment of SHACL (with minor restrictions) and then lift it to the case of *stratified SHACL* which allows both recursion and negation, but restricts their interaction.[1]

The contributions of this paper can be summarised as follows:

- In Section 3, we provide a semantics for validating SHACL in presence of ontologies, and argue why it is intuitive. For this, we introduce the notion of the austere canonical model, a canonical model which locally does not contain redundant structures, and advocate checking for validation over this specific model.

[1] The impossibility of such a rewriting for SHACL with negation given in Theorem 4.4 of [6] does not hold, neither for our semantics nor for the minimal-model semantics adopted in that work, as acknowledged by the authors in personal communication, and shown by the results of this work.

- In Section 4, we discuss that, although the austere canonical model may be infinite, we can provide a finite representation in the form of the *good successor configuration*. We show that the result of our construction coincides with the model construction proposed in [11] and is indeed universal. Moreover, we show that the local minimality of the austere canonical model suffices for global minimality: the austere canonical model is a core, and it is the unique universal core model of the consistent Horn-$\mathcal{ALCHIQ}$ TBox.
- In Section 5, we define a fragment of recursive SHACL named *stratified SHACL*. Our notion of stratification is based on the well-known class of stratified logic programs [20]. We define a *least fixed point semantics* for it that coincides with both the stable [21] and the well-founded [22] semantics. We also show that our fragment has a rather simple normal form with the same expressivity.
- In Sections 6 and 7, we are able to bring validation over the possibly infinite austere canonical model back to validation of a rewritten set of constraints over an enriched ABox. We do this by combining the normal form of stratified SHACL with the information captured in the good successor configuration.
- In Section 8, we discuss some techniques to create a pure rewriting of SHACL with ontologies into plain SHACL. One of these techniques proposes an extension of SHACL, SHACL$^b$, which also allows to define labels for roles.
- Lastly, in Section 9, we determine the complexity of validating SHACL with ontologies. We find that in presence of Horn-$\mathcal{ALCHIQ}$ TBoxes, SHACL validation is ExpTime complete in combined complexity, and PTime complete in data complexity. Moreover, we show that validating a very simple fragment of SHACL, simple SHACL, over a rather light (less expressive than *DL-Lite*$_\mathcal{R}$) description logic ontology already suffices to find ExpTime hardness in combined complexity.

## 2. Preliminaries

*Data Graphs and Interpretations.* Let $N_C$, $N_R$, $N_I$ and $N_B$ denote countably infinite, mutually disjoint sets of *concept names* (also known as *class names*), *role names* (or, *property names*), *individuals* (or, *constants*), and *blank nodes* respectively. Let $\overline{N_R} := \{p, p^- \mid p \in N_R\}$ denote *roles*, and let $\overline{N_C} := N_C \cup \{\top, \bot\}$. For every $p \in N_R$, let $(p^-)^- = p$. For each set of roles $R \subseteq \overline{N_R}$, set $R^- := \{r^- \mid r \in \overline{N_R}\}$. An *atom* (or, *assertion*) is an expression of the form $A(e)$ or $p(e, e')$, for $A \in N_C$, $p \in N_R$ and $\{e, e'\} \subseteq N_I \cup N_B$. An *ABox* (or a *data graph*) $\mathcal{A}$ is a finite set of atoms such that no blank nodes are used.

An *interpretation* is a pair $\mathcal{I} = (\Delta^\mathcal{I}, \cdot^\mathcal{I})$, where $\Delta^\mathcal{I}$ is a non-empty set (called *domain*) and $\cdot^\mathcal{I}$ is a function that maps every $A \in N_C$ to a set $A^\mathcal{I} \subseteq \Delta^\mathcal{I}$, every $p \in N_R$ to a binary relation $p^\mathcal{I} \subseteq \Delta^\mathcal{I} \times \Delta^\mathcal{I}$, and every individual and blank node $e \in N_I \cup N_B$ to an element $e^\mathcal{I} \in \Delta^\mathcal{I}$. Let $(p^-)^\mathcal{I} := \{(e', e) \mid (e, e') \in p^\mathcal{I}\}$. We make the standard name assumption, which means $e^\mathcal{I} = e$ for all interpretations $\mathcal{I}$, and all $e \in N_I \cup N_B$.

The *canonical interpretation* $\mathcal{I}_A$ for an ABox $\mathcal{A}$ is defined by setting $\Delta^{\mathcal{I}_\mathcal{A}} = N_I \cup N_B$, $A^{\mathcal{I}_\mathcal{A}} = \{c \mid A(c) \in \mathcal{A}\}$ for all $A \in N_C$, $p^{\mathcal{I}_\mathcal{A}} = \{(c, d) \mid p(c, d) \in \mathcal{A}\}$ for all $p \in N_R$, and $e^{\mathcal{I}_\mathcal{A}} = e$ for every individual or blank node $e \in N_I \cup N_B$. We consider an interpretation to be *finite* whenever $A^\mathcal{I}$ and $p^\mathcal{I}$ are finite sets, for each $A \in N_C$ and $p \in N_R$, and only finitely many concepts and roles have a non-empty interpretation.

*Morphisms.* Let $\mathcal{A}$ and $\mathcal{A}'$ be sets of atoms. A *homomorphism* from $\mathcal{A}$ to $\mathcal{A}'$ is a function $h : \Delta^{\mathcal{I}_\mathcal{A}} \to \Delta^{\mathcal{I}'_\mathcal{A}}$ such that for all $\{e, e'\} \subseteq N_I \cup N_B$, all $A \in N_C$ and all $p \in N_R$, (i) if $e \in \Delta^\mathcal{A} \cap N_I$, then $h(e) = e$, (ii) if $A(e) \in \mathcal{A}$, then $A(h(e)) \in \mathcal{A}'$, and (iii) if $p(e, e') \in \mathcal{A}$, then $p(h(e), h(e')) \in \mathcal{A}'$. A homomorphism is called *strong* if (ii) and (iii) are strengthened to "$A(e) \in \mathcal{A}$ iff $A(h(e)) \in \mathcal{A}'$" and "$p(e, e') \in \mathcal{A}$ iff $p(h(e), h(e')) \in \mathcal{A}'$", respectively. An *embedding* is a strong injective homomorphism, an *isomorphism* is a surjective embedding and an *endomorphism* of $\mathcal{A}$ is a homomorphism from $\mathcal{A}$ to itself.

*Syntax and Semantics of normalised Horn-$\mathcal{ALCHIQ}$.* Given a set $M = \{M_0, \dots, M_k\}$ consisting of only concept or role names, let $\bigsqcap M := M_0 \sqcap \dots \sqcap M_k$. $\mathcal{P}(X)$ denotes the power set of $X$. For a tuple $\vec{x} = (x_1, \dots, x_n)$ and $1 \le j \le n$, we let $\pi_i(\vec{x}) = x_i$ be its *ith projection*.

In a *normalised Horn-$\mathcal{ALCHIQ}$ TBox $\mathcal{T}$* each concept inclusion takes one of the following forms:

(F1) $A_0 \sqcap \dots \sqcap A_n \sqsubseteq B$ (F3) $A \sqsubseteq \forall r.B$

(F2) $A \sqsubseteq \le_1 r.B$ (F4) $A \sqsubseteq \exists r.B$

for $\{A, A_0, \dots, A_n, B\} \subseteq \overline{N_C}$ and $r \in \overline{N_R}$. Furthermore, $\mathcal{T}$ may contain role inclusions of the form $r \sqsubseteq r'$, for $\{r, r'\} \subseteq \overline{N_R}$. For more details and a discussion regarding going beyond normalised Horn-$\mathcal{ALCHIQ}$, we refer the reader to Section 10.

The semantics of normalised Horn-$\mathcal{ALCHIQ}$ is defined in terms of interpretations $\mathcal{I}$: a role or concept inclusion axiom $C \sqsubseteq D$ is satisfied whenever $C^\mathcal{I} \subseteq D^\mathcal{I}$. To this end, the interpretation function is extended in the following way: $\top^\mathcal{I} := \Delta^\mathcal{I}$, $\bot^\mathcal{I} := \emptyset$, $(A_0 \sqcap \dots \sqcap A_n)^\mathcal{I} := A_0^\mathcal{I} \cap \dots \cap A_n^\mathcal{I}$, $(\le_1 r.B)^\mathcal{I} := \{e \in \Delta^\mathcal{I} \mid |\{e' \in \Delta^\mathcal{I} \mid (e, e') \in r^\mathcal{I}, e' \in B^\mathcal{I}\}| \le 1\}$, $(\forall r.B)^\mathcal{I} := \{e \in \Delta^\mathcal{I} \mid \forall e' \in \Delta^\mathcal{I}.(e, e') \in r^\mathcal{I} \to e' \in B^\mathcal{I}\}$, $(\exists r.B)^\mathcal{I} := \{e \in \Delta^\mathcal{I} \mid \exists e' \in \Delta^\mathcal{I}.(e, e') \in r^\mathcal{I} \wedge e' \in B^\mathcal{I}\}$. We say a TBox $\mathcal{T}$ is satisfied in $\mathcal{I}$ if all its axioms are satisfied. In that case, we say $\mathcal{I}$ is a model of $\mathcal{T}$. To denote logical entailment, we may write $\mathcal{T} \models \gamma$ if every model of the TBox $\mathcal{T}$ is also a model of $\gamma$ (where the latter may be any inclusion, a TBox, or an ABox). We call the combination of a Horn-$\mathcal{ALCHIQ}$ TBox $\mathcal{T}$ and any ABox $\mathcal{A}$ a Horn-$\mathcal{ALCHIQ}$ knowledge base $(\mathcal{T}, \mathcal{A})$. We say that $\mathcal{A}$ is *consistent* with $\mathcal{T}$ (or, that $(\mathcal{T}, \mathcal{A})$ is consistent) if there is a model of $\mathcal{A}$ and $\mathcal{T}$.

We call an interpretation $\mathcal{I}$ a *universal model* (of a knowledge base $(\mathcal{T}, \mathcal{A})$) whenever $\mathcal{I}$ is a model of $(\mathcal{T}, \mathcal{A})$ and there exists a homomorphism of $\mathcal{I}$ into any model of $(\mathcal{T}, \mathcal{A})$. Every consistent $(\mathcal{T}, \mathcal{A})$ has a universal model [23]. Moreover, the universal model of $(\mathcal{T}, \mathcal{A})$ coincides with the universal model of $(\mathcal{T}^{pos}, \mathcal{A})$, where $\mathcal{T}^{pos}$ contains all axioms in $\mathcal{T}$ that do not contain $\bot$. Therefore, we may assume that there are no occurrences of $\bot$ if $(\mathcal{T}, \mathcal{A})$ is consistent.

$$\begin{aligned}
\mathcal{I}(c) &:= \{c^{\mathcal{I}}\}\\
\mathcal{I}(A) &:= A^{\mathcal{I}}\\
\mathcal{I}(\neg\varphi) &:= \Delta^{\mathcal{I}} \setminus \mathcal{I}(\varphi)\\
\mathcal{I}(\varphi \wedge \varphi') &:= \mathcal{I}(\varphi) \cap \mathcal{I}(\varphi')\\
\mathcal{I}(\varphi \vee \varphi') &:= \mathcal{I}(\varphi) \cup \mathcal{I}(\varphi')\\
\mathcal{I}(\exists_{\geq n} E.\varphi) &:= \{e \in \Delta^{\mathcal{I}} \mid |\{e' \in \Delta^{\mathcal{I}} \mid (e,e') \in E^{\mathcal{I}} \wedge e' \in \mathcal{I}(\varphi)\}| \geq n\}\\
\mathcal{I}(\mathsf{eq}(E,r)) &:= \{e \in \Delta^{\mathcal{I}} \mid \{\exists e' \in \Delta^{\mathcal{I}} \mid (e,e') \in E^{\mathcal{I}}\} = \{e' \in \Delta^{\mathcal{I}} \mid (e,e') \in r^{\mathcal{I}}\}\}\\
\mathcal{I}(\mathsf{disj}(E,r)) &:= \{e \in \Delta^{\mathcal{I}} \mid \{\exists e' \in \Delta^{\mathcal{I}} \mid (e,e') \in E^{\mathcal{I}}\} \cap \{e' \in \Delta^{\mathcal{I}} \mid (e,e') \in r^{\mathcal{I}}\} = \emptyset\}\\
\mathcal{I}(\mathsf{closed}(R)) &:= \{e \in \Delta^{\mathcal{I}} \mid \text{ there is no } r \in \overline{N_R} \setminus R \text{ such that } e \in (\exists r.\top)^{\mathcal{I}}\}
\end{aligned}$$

**Fig. 1.** Evaluating shape expressions.

*Regular Path Expressions.* Let $E$ be any *regular expression* over some alphabet $\Sigma$, and $L_E$ the language defined by some regular expression $E$. We say $E$ is a regular *path* expression, if $\Sigma = \overline{N_R}$. For each interpretation $\mathcal{I}$ and each regular path expression $E$ over the alphabet $\overline{N_R}$, set $(e, e') \in E^{\mathcal{I}}$ if there exists $r_0 \cdots r_n \in L_E$ and $\{e_1, \ldots, e_n\} \subseteq \Delta^{\mathcal{I}}$ such that $(e, e_1) \in r_0^{\mathcal{I}}$, $(e_n, e') \in r_n^{\mathcal{I}}$ and for all $1 \leq i \leq n-1$, $(e_i, e_{i+1}) \in r_i^{\mathcal{I}}$.

Furthermore, for each language $L_E$, there exists a non-deterministic finite automaton $\mathcal{M} = (Q, \Sigma, q_I, \Delta, q_F)$ that accepts exactly all words in $L_E$. Here $Q$ is a set of states, $\Sigma$ the alphabet, $\{q_I, q_F\} \subseteq Q$ the initial and final state, and $\Delta \subseteq Q \times \Sigma \times Q$ the transition relation. In this case, we say $\mathcal{M}$ recognises $L_E$.

*Non-Recursive Shape Constraint Language (SHACL).* We define *shape expressions*, following [24], in the following way

$$\varphi ::= c \mid A \mid \top \mid \neg\varphi \mid \varphi \wedge \varphi \mid \varphi \vee \varphi \mid \exists_{\geq n} E.\varphi \mid \mathsf{eq}(E, r) \mid \mathsf{disj}(E, r) \mid \mathsf{closed}(R),$$

where $c \in N_I$, $A \in N_C$, $n \geq 1$, $R$ a finite subset of $\overline{N_R}$ and $E$ a regular path expression. We use $\top$ as a shorthand for $A \vee \neg A$, for some concept name $A \in N_C$. As we impose restrictions on the use of negation later in this work, we add $\varphi \vee \varphi$ and $\varphi \wedge \varphi$ explicitly to the syntax.

Given an interpretation $\mathcal{I}$, we say a node $e \in N_I \cup N_B$ *validates* a shape expression $\varphi$, if $e \in \mathcal{I}(\varphi)$, where $\mathcal{I}(\varphi)$ is inductively defined in Fig. 1. Furthermore, let $\mathcal{G}$ be a set of targets of the form $\varphi(c)$ or $\varphi(A)$, where $\varphi$ is a shape expression, $c \in N_I$ and $A \in N_C$. Given an interpretation $\mathcal{I}$, we say $\mathcal{I}$ *validates* $\mathcal{G}$ if for all $\varphi(c) \in \mathcal{G}$, we find $c$ validates $\varphi$, and for all $\varphi(A) \in \mathcal{G}$, if $c \in A^{\mathcal{I}}$, then $c$ validates $\varphi$. Considering readability, we will also write $\mathcal{A}$ validates $\mathcal{G}$, for any set of atoms $\mathcal{A}$, instead of using the canonical interpretation $\mathcal{I}_{\mathcal{A}}$.

In this definition, we do not allow shape names in shape expressions, as introduced in Section 5. This makes SHACL *non-recursive* and allows for the simple semantics we define above. In general, SHACL allows shape names in shapes expressions and may be *recursive*, in which case there is no unique accepted semantics [21,22,25]. For simplicity, we formulate all the results in Sections 3 and 4 for non-recursive SHACL. However, the results and definitions can be directly applied to recursive SHACL under all the semantics considered in [21,22,25]. Starting from Section 5, we concentrate on a recursive form of SHACL, stratified SHACL, extending the work presented in [15]. The fragment we study in this work is not the full stratified, recursive version; it lacks counting quantifiers and the closed-operator, and only treats a guarded version of disjointness and equality. This is discussed in more detail in Sections 5 and 10. Removing these constraints leads to new interesting problems we leave for future work.

## 3. Validation with ontologies

In this section we propose a semantics for SHACL validation in the presence of a Horn-$\mathcal{ALCHIQ}$ ontology. More precisely, for a given TBox $\mathcal{T}$, an ABox $\mathcal{A}$, and a set of targets $\mathcal{G}$, we aim to define when $(\mathcal{T}, \mathcal{A})$ validates $\mathcal{G}$. A natural first idea would be to follow the usual open-world semantics of Horn-$\mathcal{ALCHIQ}$ and check validation on *all models* of $\mathcal{A}$ and $\mathcal{T}$. While this works well for positive constraints, it does not yield a natural semantics in the presence of negation, as illustrated in the following simple example.

**Example 3.1.** Consider the ABox $\mathcal{A}$ consisting of the facts $hasPet(linda, blu)$, $Bird(blu)$, an empty TBox $\mathcal{T}$ and the target $\varphi(linda)$ such that $\varphi$ is given by

$$\varphi = \exists hasPet.\neg Dog.$$

Naturally, $\mathcal{A}$ validates $\mathcal{G}$, as $linda$ indeed validates $\varphi$, which corresponds to having a pet that is not a dog. Note that since we have an empty TBox, we would like to be in the usual setting of validation here. That is, one would expect $(\mathcal{T}, \mathcal{A})$ to validate $\mathcal{G}$. However, if we consider all possible models of $(\mathcal{T}, \mathcal{A})$, we find non-validation: the case in which $blu$ is both a $Bird$ and a $Dog$ is also a model of $(\mathcal{T}, \mathcal{A})$. This shows that the certain answer semantics, guaranteeing validation in all models, is not always meaningful in the presence of negation. ♡

The above example illustrates the problem of finding an intuitive semantics for shape expressions, or, on the same note, queries, with negation. Similar issues show up in shape expressions containing a more hidden form of negation, like equality, disjointness and the closed-operator. Also counting quantifiers have the capability of distinguishing between different universal models. Roughly speaking, adding facts to the data may cause a previously validated setting to become invalid. To this end, we aim at an intuitive semantics that coincides with the usual validation in case the TBox is empty, but also lets the TBox axioms influence the validation results in the relevant cases. As done in related settings (see e.g., [14,18,26]) we rely on the *chase* procedure [27] known from Knowledge Representation and Database Theory. The idea is that a chase procedure takes as input an ABox and TBox and iteratively applies the axioms of the TBox to the data by adding atoms over possibly fresh individuals until all the axioms in the TBox are satisfied. The result of the chase is a so-called *canonical* or *universal* model. Since there exists a homomorphism of all models of this type into every other model of the ABox and TBox, a canonical model is often used as a representative of all models.

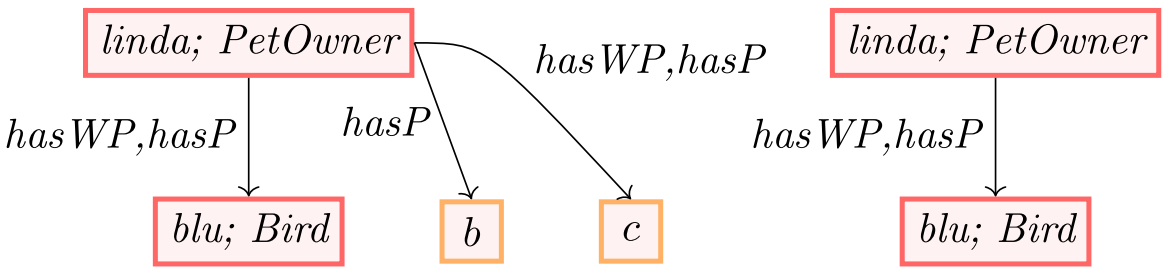

**Fig. 2.** Result of oblivious or Skolem chase (left) and austere canonical model (right). We use hasWP and hasP as a shorthand for hasWingedPet and hasPet.

There are several chase variants producing canonical models with different properties [26]. While for positive shape expressions these differences do not lead to different validation results, shape expressions using negation can distinguish between them. Thus, the semantics we propose is based on a particular variant of the set of universal models, based on local minimality. The idea is to avoid redundant structures as much as we can: as illustrated in the next example, we do not wish to assume the existence of a second pet of $linda$ if there is no need for this assumption. At the same time, we do not want to give up on the model being universal. This specific model, the *austere canonical model*, will be constructed in the rest of this section.

In the following example, we show that shape expressions with negation can indeed distinguish between different universal models, and illustrate how the austere canonical model does not have the redundant structures that may appear in other universal models.

**Example 3.2.** Consider the ABox $\mathcal{A} = \{PetOwner(linda), hasWingedPet(linda, blu), Bird(blu)\}$ and the following three axioms:

$$PetOwner \sqsubseteq \exists hasPet, \qquad hasWingedPet \sqsubseteq hasPet,$$
$$PetOwner \sqsubseteq \exists hasWingedPet.$$

The austere canonical model (right in Fig. 2) will only add a $hasPet$-role from $linda$ to $blue$, as we will see below. In contrast, the canonical model obtained from the oblivious chase or Skolem chase (left in Fig. 2) will introduce two fresh objects to satisfy the two axioms with existential restrictions.

When graphically representing interpretations, domain objects are written in rectangular boxes (individuals in red and blank nodes in orange boxes), followed by a semicolon and the concept names in whose interpretations it participates, if any. Roles are depicted as labelled arrows.

Now let us consider the same shape expression and target as in the previous example: $\varphi = \exists hasPet.\neg Bird\}$ and $\mathcal{G} = \{\varphi(linda)\}$. The target asks to validate whether $linda$ has a pet that is not a bird. Clearly, the austere canonical model provides the expected answer, as it does not validate $\mathcal{G}$. In contrast, the canonical model on the left-hand-side of the figure, also the semantics of [14] adopted for SHACL in [6], results in the unintended validation of $(\mathcal{C}, \mathcal{G})$. ♡

In the rest of this section, we will make the above precise.

### 3.1. Good successor configuration

To capture local minimality, we define the auxiliary notion of the *good successor configuration*. It determines, for each point $e$ in a model, the set of fresh successor individuals and the roles connecting them to $e$ in the austere canonical model, together with the concepts that must hold at $e$. To describe the good successor configuration, we use *2-types*, *1½-types* and *1-types*. We say $t$ is a 2-type if $t \in \mathcal{P}(N_C) \times \mathcal{P}(\overline{N_R}) \times \mathcal{P}(N_C)$. Similarly, we let $u$ be a *1½*-type in case $u \in \mathcal{P}(\overline{N_R}) \times \mathcal{P}(N_C)$. A 2-type describes a pair of nodes and the roles between them, while a *1½*-type describes a set of roles leading to one node. A *1*-type $c \in \mathcal{P}(N_C)$ is simply a set of concept names. Furthermore, we define the *inverse* function $inv$ mapping a 2-type to a 2-type by setting $inv(t) := (\pi_3(t), (\pi_2(t))^-, \pi_1(t))$.

To understand the good successor configuration, assume we are building a model in a chase-like step-by-step manner. We want to introduce the children that a node $c$ needs to satisfy the TBox. Let $c_1, \dots, c_n$ be the neighbours that $c$ already has (either multiple neighbours already given in the ABox, or the one single neighbour/parent $c_1$ has in the anonymous part). We use a set $F$ of 2-types to describe $c$ and its environment as follows. For each $c_i$, some type $t_i \in F$ describes $c$ in relation to $c_i$: (i) $\pi_1(t_i)$ is the *1*-type of $c$, that is, the concept names whose interpretations contain $c$, (ii) $\pi_3(t_i)$ is the *1*-type of $c_i$, and (iii) $\pi_2(t_i)$ contains all role names connecting $c$ and $c_i$. Note that, by item (i), $\pi_1(t_i) = \pi_1(t_j)$ for all $\{t_i, t_j\} \subseteq F$. The good successor configuration $succ_\mathcal{T}$, which is defined in the following definition, is a function that takes such an $F$, and returns the description of the children that $c$ needs to satisfy all axioms in $\mathcal{T}$, as a set of *1½*-types: we will add a new blank node $d_i$ for each $u_i \in succ_\mathcal{T}(F)$; $d_i$ will be connected to $c$ via the roles in $\pi_1(u_i)$, and have *1*-type $\pi_2(u_i)$.

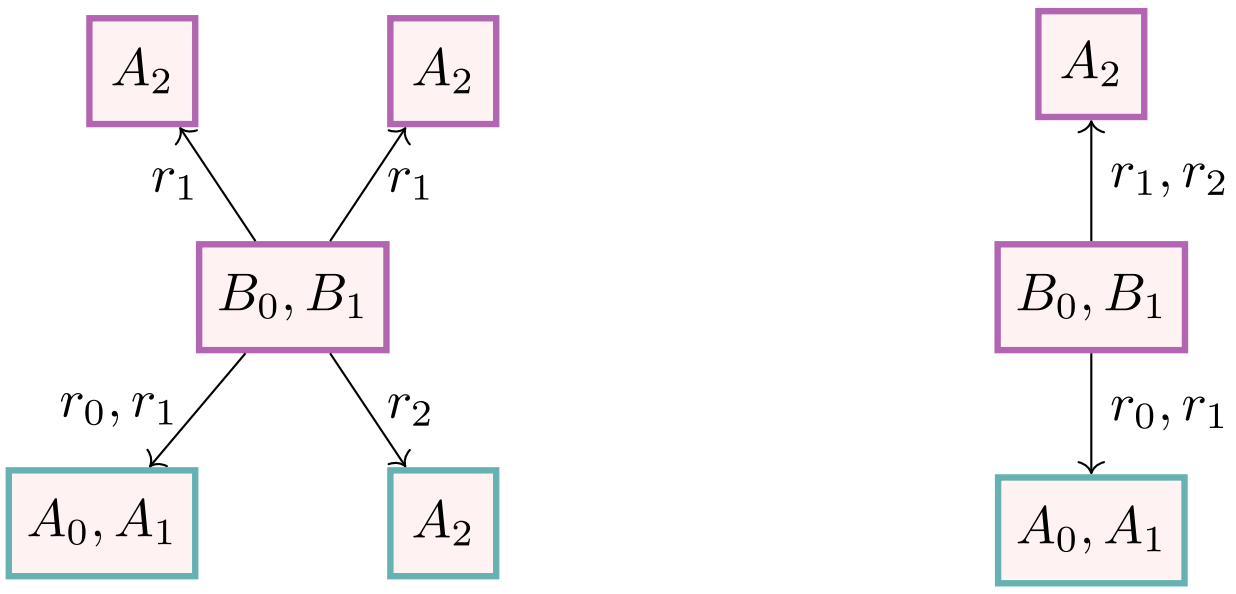

**Fig. 3.** Given the TBox $\mathcal{T}$ as in Example 3.4, the good successor configurations for the sets of 2-types $F = \{(\{B_0, B_1\}, \{r_1\}, \{A_2\}), (\{B_0, B_1\}, \{r_1\}, \{A_2\})\}$ and $F' = \{(\{B_0, B_1\}, \{r_1, r_2\}, \{A_2\})\}$, are $succ_{\mathcal{T}}(F) = \{(\{r_0, r_1\}, \{A_0, A_1\}), (\{r_2\}, \{A_2\})\}$ and $succ_{\mathcal{T}}(F') = \{(\{r_0, r_1\}, \{A_0, A_1\})\}$.

However, the good successor configuration does not just take specific neighbourhoods of the nodes in some database as input; it is defined for all possible $F$'s. This makes the good successor configuration data independent. Thus, when determining the data complexity of any algorithm using the good successor configuration, its size does not have to be considered.

Moreover, the following definition defines a unique set of 1½-types; this is proven at the end of this subsection. Finally, note that we use expressions of the form $\exists(\bigsqcap R).\bigsqcap N$ below, for $R \subseteq \overline{N_R}$, $N \subseteq N_C$ with the following semantics:

$$(\exists(\bigsqcap R).\bigsqcap N)^{\mathcal{I}} := \{e \in \Delta^{\mathcal{I}} \mid \exists e' \in \Delta^{\mathcal{I}} \forall r \in R.(e, e') \in r^{\mathcal{I}} \wedge e' \in (\bigsqcap N)^{\mathcal{I}}\}.$$

**Definition 3.3.** Given a normalised Horn-$\mathcal{ALCHIQ}$ TBox $\mathcal{T}$ and set of 2-types $F$ such that $\pi_1(t) = \pi_1(t')$ for all $\{t, t'\} \subseteq F$, the *good successor configuration* $succ_{\mathcal{T}}(F)$ is a possibly empty set of 1½-types $u$ such that:

(R1) If (i) $M \subseteq \pi_1(t)$ for some $t \in F$, (ii) $\mathcal{T} \models \bigsqcap M \sqsubseteq \exists(\bigsqcap R).\bigsqcap N$ for some $N \subseteq N_C$, $R \subseteq \overline{N_R}$, and (iii) for all $t \in F$, $R \not\subseteq \pi_2(t)$ or $N \not\subseteq \pi_3(t)$, then there exists $u \in succ_{\mathcal{T}}(F)$ such that $R \subseteq \pi_1(u)$ and $N \subseteq \pi_2(u)$;
(R2) If $u \in succ_{\mathcal{T}}(F)$, then there exist $M \subseteq \pi_1(t)$ for some $t \in F$, such that $\mathcal{T} \models \bigsqcap M \sqsubseteq \exists(\bigsqcap \pi_1(u)).\bigsqcap \pi_2(u)$;
(R3) There does not exist $\{u, u'\} \subseteq succ_{\mathcal{T}}(F)$ with $u \neq u'$ such that $\pi_1(u) \subseteq \pi_1(u')$ and $\pi_2(u) \subseteq \pi_2(u')$;
(R4) If $u \in succ_{\mathcal{T}}(F)$, then $\pi_1(u) \not\subseteq \pi_2(t)$ or $\pi_2(u) \not\subseteq \pi_3(t)$, for all $t \in F$.

A *child* of a 2-type $t$ is a 2-type $t' \in child_{\mathcal{T}}(t)$ such that $\pi_1(t') = \pi_3(t)$ and $(\pi_2(t'), \pi_3(t')) \in succ_{\mathcal{T}}(inv(t))$.

For readability, we define the good successor configuration of a single type as the good successor configuration of the set containing only this type.

As mentioned, the good successor configuration focusses on checking which axioms are not yet satisfied in the context described by $F$. For each such unsatisfied axiom (R1) implies the existence of a child $d_i$ represented by some $u_i$ that ensures the satisfaction of the axiom. (R2) represents the other direction: for each child $d_i$, there must exist an axiom implying the information in the 1½-type $u_i$. Lastly, (R3) and (R4) check whether there are no superfluous children: they enforce we do not add two children $d_i$ and $d_j$ such that all information about $d_i$ is subsumed by $d_j$, or a child $d_i$ that is already subsumed by the environment described in $F$. We will use these properties extensively in Section 5 to prove that the austere canonical model is in fact a core.

In the following example, we compute the good successor configuration for a concrete case.

**Example 3.4.** Consider the axioms below, together forming the TBox $\mathcal{T}$.

(T1) $B_0 \sqsubseteq \exists r_0.A_0$
(T2) $B_0 \sqsubseteq \exists r_1.A_1$
(T3) $B_1 \sqsubseteq \leq_1 r_1.A_1$
(T4) $A_0 \sqsubseteq A_1$
(T5) $r_0 \sqsubseteq r_1$
(T6) $B_1 \sqsubseteq \exists r_2.\top$
(T7) $B_0 \sqsubseteq \forall r_2.A_2$

We want to compute the good successor configuration for the following sets of 2-types: $F = \{(\{B_0, B_1\}, \{r_1\}, \{A_2\}), (\{B_0, B_1\}, \{r_1\}, \{A_2\})\}$ and $F' = \{(\{B_0, B_1\}, \{r_1, r_2\}, \{A_2\})\}$.

First, note that (T1–5) together lets us conclude $\mathcal{T} \models B_0 \sqcap B_1 \sqsubseteq \exists(r_0 \sqcap r_1).(A_0 \sqcap A_1)$. Moreover, combining (T6) and (T7) gives us $\mathcal{T} \models B_0 \sqcap B_1 \sqsubseteq \exists r_2.A_2$. Thus, by (R1), we find that there must exist $\{u, u'\} \subseteq succ_{\mathcal{T}}(F)$ such that $\{r_0, r_1\} \subseteq \pi_1(u)$, $\{A_0, A_1\} \subseteq \pi_2(u)$, $\{r_2\} \subseteq \pi_1(u')$ and $\{A_2\} \subseteq \pi_2(u')$. Note that considering $\pi_1(u) = \{r_0, r_1\}$, $\pi_2(u) = \{A_0, A_1\}$, $\pi_1(u') = \{r_2\}$ and $\pi_2(u') = \{A_2\}$ suffices, as we know that these are the axioms carrying most information that can be derived. It is easy to check that indeed $succ_{\mathcal{T}}(F) = \{u, u'\}$; see also Fig. 3.

For computing the good successor configuration of $F'$, we can mostly follow the above strategy. However, $u'$ cannot be added by (R1), as $u'$ is already contained in the environment described by $F'$. This does not hold for $u$, so it is easy to check that $succ_{\mathcal{T}}(F') = \{u\}$. ♡

Essential to the rest of this article is the uniqueness and existence of the good successor configuration. This implies that the construction described in the next chapter creates a unique structure.

**Proposition 3.5.** *Given a normalised Horn-$\mathcal{ALCHIQ}$ TBox $\mathcal{T}$, for each set of 2-types $F$, such that $\pi_1(t) = \pi_1(t')$ for all $\{t, t'\} \subseteq F$, there exists a unique good successor configuration $succ_{\mathcal{T}}(F)$.*

**Proof.** To show uniqueness, suppose for a given set of 2-types $F$, such that $\pi_1(t) = \pi_1(t')$ for all $\{t, t'\} \subseteq F$, there exist two good successor configurations $U \neq U'$. W.l.o.g., assume that there exists $u \in U$, such that $u \notin U'$. By (R2), we know that there exists $M \subseteq \pi_1(t)$ for some $t \in F$ such that $\mathcal{T} \models \bigsqcap M \sqsubseteq \exists(\bigsqcap \pi_1(u)).\bigsqcap \pi_2(u)$. Moreover, by (R4), we can conclude that $\pi_1(u) \nsubseteq \pi_2(t)$ or $\pi_2(u) \nsubseteq \pi_3(t)$ for all $t \in F$, so we can apply rule (R1) with respect to $U'$: there must exist $u' \in U'$ such that $\pi_1(u) \subseteq \pi_1(u')$ and $\pi_2(u) \subseteq \pi_2(u')$. By similar reasoning, we find that there must exist $u'' \in U$ such that $\pi_1(u') \subseteq \pi_1(u'')$ and $\pi_2(u') \subseteq \pi_2(u'')$. Combining this with (R4), we find that $\pi_1(u) = \pi_1(u'')$ and $\pi_2(u) = \pi_2(u'')$, from which we derive that $u \in U'$, which is a contradiction. Thus, $U = U'$. For existence, note that we can first add all possible *1½*-types that are not eliminated by (R2). It follows that (R1) is clearly satisfied. Then, remove all *1½*-types that are eliminated because of (R3) or (R4). ♣

By using the inference rules from Table 2 in [11], combined with the approach described in this proof, it becomes clear that the good successor configuration can actually be computed.

### *3.2. Austere canonical model*

The good successor configuration locally describes how to satisfy the TBox axioms based on some incentives; minimality and universality. Before we use it as a building block in the austere canonical model, we first want to complete $\mathcal{A}$ under all but the axioms with existential restrictions. We will use the notation $\mathcal{A}_{\mathcal{T}}$ for this completed ABox, which is in fact the least fixed point of an immediate consequence operator, a monotone function. Note that a function $f$ is *monotone* if $x \subseteq y$ implies $f(x) \subseteq f(y)$. The point of this operator is to mimic firing the axioms without existential restrictions as soon as they become applicable. We perform the least fixed point starting from a non-empty set: the ABox. This is made precise in the following definitions.

**Definition 3.6.** Let $X$ be any set and $T : X \to X$ be any monotone function, which we will call an *immediate consequence operator*, define for any $S \subseteq X$, $T \uparrow^{\omega} (S)$ as follows:

1. $T \uparrow^0 (S) := S$,
2. $T \uparrow^{n+1} (S) := T(T \uparrow^n (S))$ for $n \geq 0$, and
3. $T \uparrow^{\omega} (S) := \bigcup_{n=0}^{\infty} T \uparrow^n (S)$.

**Definition 3.7** (Completion $\mathcal{A}_{\mathcal{T}}$). Given a normalised Horn-$\mathcal{ALCHIQ}$ TBox $\mathcal{T}$, define an immediate consequence operator $T_{\mathcal{T}}$ that maps a set of atoms $X$ to a set of atoms as follows:

$$\begin{aligned}
T_{\mathcal{T}}(X) := X &\cup \{B(a) \mid \{A_0(a), \ldots, A_n(a)\} \subseteq X, \mathcal{T} \models A_0 \sqcap \ldots \sqcap A_n \sqsubseteq B\} \\
&\cup \{B(a) \mid \{r(b,a), A(b)\} \subseteq X, A \sqsubseteq \forall r.B \in \mathcal{T}\} \\
&\cup \{r(a,b) \mid r'(a,b) \in X, \mathcal{T} \models r' \sqsubseteq r\} \\
&\cup \{r(b,a) \mid r'(a,b) \in X, \mathcal{T} \models r' \sqsubseteq r^-\} \\
&\cup \{B_i(b) \mid \{A(a), A_1(a), \ldots, A_n(a), r(a,b), B(b)\} \subseteq X, A \sqsubseteq \leq_1 r.B \in \mathcal{T}, \\
&\qquad \mathcal{T} \models A_1 \sqcap \ldots \sqcap A_n \sqsubseteq \exists(r_1 \sqcap \ldots \sqcap r_m).B_1 \sqcap \ldots \sqcap B_m, A = A_j, r = r_k \text{ for some } j,k\} \\
&\cup \{r_i(a,b) \mid \{A(a), A_1(a), \ldots, A_n(a), r(a,b), B(b)\} \subseteq X, A \sqsubseteq \leq_1 r.B \in \mathcal{T}, \\
&\qquad \mathcal{T} \models A_1 \sqcap \ldots \sqcap A_n \sqsubseteq \exists(r_1 \sqcap \ldots \sqcap r_m).B_1 \sqcap \ldots \sqcap B_m, A = A_j, r = r_k \text{ for some } j,k\}.
\end{aligned}$$

Given any ABox $\mathcal{A}$, let $N_I(\mathcal{A})$ be the set of individuals occurring in $\mathcal{A}$. We set $A(a) \in \mathcal{A}_{\mathcal{T}}$ and $r(a,b) \in \mathcal{A}_{\mathcal{T}}$, iff $A(a) \in T_{\mathcal{T}} \uparrow^{\omega} (\mathcal{A})$, respectively $r(a,b) \in T_{\mathcal{T}} \uparrow^{\omega} (\mathcal{A})$, for all $A \in N_C$, $r \in N_R$, $\{a,b\} \subseteq N_I(\mathcal{A})$.

It can be verified how the rules in this definition mimic the completion rules for ABoxes under Horn-$\mathcal{SHIQ}$ ontologies in [11].

**Example 3.8.** Let $\mathcal{A} = \{B_0(a), r_0(a,b), r_2(a,b), A_0(b)\}$ and $\mathcal{T}$ as in Example 3.4. Then $\mathcal{A}_{\mathcal{T}} = \mathcal{A} \cup \{r_1(a,b), A_1(b), A_2(b)\}$. ♡

Now we are ready to define the austere canonical model. In this model, we use words to represent the elements in the anonymous parts of the model: the first letter corresponds to an individual in the data graph, followed by a letter for each node connecting the named node to the data graph. Thus, the length of the word naming an anonymous node indicates its 'distance' from the data graph. We mainly use this distance in the next section in which we discuss the relation to core models.

**Definition 3.9.** Let $(\mathcal{T}, \mathcal{A})$ be a normalised Horn-$\mathcal{ALCHIQ}$ knowledge base, and let $N_{(\mathcal{T},\mathcal{A})} \subseteq N_B$ be the set of finite words of the form $ak_1 \ldots k_n$, with $a \in N_I(\mathcal{A})$ such that for all $1 \leq i \leq n$, $k_i$ is a 2-type such that the following hold:

1. $\pi_1(k_1) = \{A \mid A(a) \in \mathcal{A}_{\mathcal{T}}\}$ and $(\pi_2(k_1), \pi_3(k_1)) \in succ_{\mathcal{T}}(T_a)$, for
$$T_a = \{(\{A \mid A(a) \in \mathcal{A}_{\mathcal{T}}\}, \emptyset, \emptyset)\} \cup \bigcup_{b \in N_I, r(a,b) \in \mathcal{A}_{\mathcal{T}}} \{(\{A \mid A(a) \in \mathcal{A}_{\mathcal{T}}\}, \{r \mid r(a,b) \in \mathcal{A}_{\mathcal{T}}\}, \{A \mid A(b) \in \mathcal{A}_{\mathcal{T}}\})\};$$
2. for every $1 \leq i < n$, $k_{i+1} \in child_{\mathcal{T}}(k_i)$.

We use $tail(w)$ to denote the last 2-type $k_n$ in, and $|w| = n + 1$ as the length of a word $w \in N_{(\mathcal{T},\mathcal{A})}$ of the form $ak_1 \ldots k_n$.

The *austere canonical model* $can(\mathcal{T},\mathcal{A})$ of a normalised Horn-$\mathcal{ALCHIQ}$ knowledge base $(\mathcal{T},\mathcal{A})$ is the interpretation $can(\mathcal{T},\mathcal{A})$ with domain $\Delta^{can(\mathcal{T},\mathcal{A})} := N_I(\mathcal{A}) \cup N_{(\mathcal{T},\mathcal{A})}$ such that for all $a \in N_I(\mathcal{A})$, concept names $A$ and roles $r$, the following hold:

$$\begin{aligned} a^{can(\mathcal{T},\mathcal{A})} &:= a \\ A^{can(\mathcal{T},\mathcal{A})} &:= \{a \in N_I \mid A(a) \in \mathcal{A}_\mathcal{T}\} \cup \{w \in N_{(\mathcal{T},\mathcal{A})} \mid A \in \pi_3(tail(w))\} \\ r^{can(\mathcal{T},\mathcal{A})} &:= \{(a,b) \in N_I \times N_I \mid r(a,b) \in \mathcal{A}_\mathcal{T}\} \cup \{(w_1,w_2) \in (N_I \cup N_{(\mathcal{T},\mathcal{A})}) \times N_{(\mathcal{T},\mathcal{A})} \mid w_2 = w_1 k, r \in \pi_2(k)\} \cup \\ &\quad\ \{(w_2,w_1) \in (N_I \cup N_{(\mathcal{T},\mathcal{A})}) \times N_{(\mathcal{T},\mathcal{A})} \mid w_2 = w_1 k, r^- \in \pi_2(k)\}. \end{aligned}$$

This is illustrated in the following example.

**Example 3.10.** Let $\mathcal{A} = \{B_0(a), B_1(a), r_1(a,b), r_1(a,c), A_2(b), A_2(c)\}$ and suppose $\mathcal{T}$ contains the same axioms as in Example 3.4. We find that $\mathcal{A}_\mathcal{T} = \mathcal{A}$, as none of the axioms there can derive any new information about ABox individuals.

Now consider the set $F_a = \{(\{B_0,B_1\},\{r_1\},\{A_2\}),(\{B_0,B_1\},\{r_1\},\{A_2\})\}$, based on the neighbours of $a$. The good successor configuration of this set has already been computed in 3.4: $succ_\mathcal{T}(F_a) = \{u,u'\}$, where $u = (\{r_0,r_1\},\{A_0,A_1\})$ and $u' = (\{r_2\},\{A_2\})$. Following the definition of the austere canonical model, we find $\{at, at'\} \subseteq N_{(\mathcal{T},\mathcal{A})}$, for $t = (\{B_0,B_1\},\pi_1(u),\pi_2(u))$ and $t' = (\{B_0,B_1\},\pi_1(u'),\pi_2(u'))$. Now we can read off the structure of the anonymous part of the constructed model: we find for instance that $(a,at) \in r_0^{can(\mathcal{T},\mathcal{A})}$ and $at \in A_0^{can(\mathcal{T},\mathcal{A})}$. ♡

For a consistent knowledge base $(\mathcal{T},\mathcal{A})$, the austere canonical model $can(\mathcal{T},\mathcal{A})$ exists and is unique, which follows directly from the uniqueness and existence of the good successor configuration and $\mathcal{A}_\mathcal{T}$.

Note that the austere canonical model coincides with the result of the model construction presented in [11]. In that construction, $\mathcal{A}$ is first closed[2] under all implied axioms, except for the ones with existential restrictions, followed by applying all so-called 'maximal' axioms with existential restrictions in a chase like manner whenever applicable. To see the correspondence with our construction, it suffices to check that Definition 3.7 formalises the first step, whilst the good successor configuration captures exactly the 'maximality' of axioms in (R3) and the applicability in (R3) and (R4).

**Corollary 3.11.** *For each normalised Horn-$\mathcal{ALCHIQ}$ knowledge base $(\mathcal{T},\mathcal{A})$, $can(\mathcal{T},\mathcal{A})$ is (i) a model of $(\mathcal{T},\mathcal{A})$ that is (ii) universal.*

This result follows directly from Proposition 5.4 in [11] and justifies the name 'canonical' we gave to our construction. Finally, we define SHACL validation with Horn-$\mathcal{ALCHIQ}$ as validation of SHACL over the austere canonical model.

**Definition 3.12** (Validation with Horn-$\mathcal{ALCHIQ}$). Given a normalised Horn-$\mathcal{ALCHIQ}$ knowledge base $(\mathcal{T},\mathcal{A})$ and a set of targets $\mathcal{G}$, we say $(\mathcal{T},\mathcal{A})$ validates $\mathcal{G}$ if $can(\mathcal{T},\mathcal{A})$ validates $\mathcal{G}$.

In general, given any semantics for SHACL validation, we define the semantics of SHACL validation with Horn-$\mathcal{ALCHIQ}$ ontologies, as validation of SHACL over the austere canonical model. In particular, this approach will be used for stratified SHACL, a fragment of recursive SHACL. We refer the reader to Section 5 for more details.

## 4. Finite and infinite cores

In database theory, the property of an interpretation or structure being a core is well studied. It is a property that ultimately represents the lack of redundant structures. Therefore, this property is not only nice to have, but particularly relevant in our setting. In fact, we will show that the austere canonical model is a core. Before we get there, we first provide the required theoretical background.

To start, we say an interpretation $\mathcal{I}$ is a *core* if each endomorphism of $\mathcal{I}$ is an embedding. That is, each homomorphism of $\mathcal{I}$ into itself is both strong and injective. The core of a set of atoms $\mathcal{A}$ is a set of atoms $\mathcal{B} \subseteq \mathcal{A}$, such that (i) there exists an endomorphism $h$ from $\mathcal{A}$ to $\mathcal{B}$, (ii) $\mathcal{B}$ is the restriction to the image of $h$, and (iii) $\mathcal{B}$ is a core. We write $\mathcal{A} \xrightarrow{core} \mathcal{B}$. Each finite set of atoms has a core that is unique up to isomorphism [28].

**Example 4.1.** Recall Fig. 2, which illustrates the results of the oblivious chase and the austere canonical model of some given knowledge base. The oblivious chase model is the following set of facts:

$$X = \{PetOwner(linda), Bird(blu), hasWingedPet(linda,blu), hasPet(linda,blu), \\ hasWingedPet(linda,b), hasPet(linda,b), hasPet(linda,c)\}$$

where $b$ and $c$ are unknown individuals, i.e., elements of the set $N_B$, whereas $\{linda, blu\} \subseteq N_I$. Thus, in each homomorphism from $X$ to $X$, that is, each endomorphism of $X$, '$linda$', and '$blu$' must be mapped to themselves, which is not true for $b$ and $c$. It is easy to check that $h : X \to X$ given by $h(linda) = linda$, $h(blu) = blu$, $h(b) = blu$ and $h(c) = blu$ is an homomorphism, but not injective or strong. Thus, $h$ is not an embedding and $X$ is not a core. Nevertheless, the austere canonical model described in the same Fig. 2, is a core, and also the core of $X$. ♡

[2] The conclusion of rule $\mathbf{R}_\leq$ should be updated to $M \sqcap M' \sqcap A \sqsubseteq \exists(S \sqcap S' \sqcap r).(N \sqcap N' \sqcap B)$, as confirmed by the authors of [11].

In the rest of this section, we first show that the austere canonical model is the unique universal core model of a given normalised Horn-$\mathcal{ALCHIQ}$ knowledge base $(\mathcal{T}, \mathcal{A})$. After that, we discuss the tight connection between the austere canonical model and the core chase.

In the rest of this section, we also allow sets of atoms to be models. This is a shorthand for saying that the canonical interpretation of this set is a model.

### 4.1. Universal core model

To show that $can(\mathcal{T}, \mathcal{A})$ is a core, we use the notion *core cover*: an interpretation $\mathcal{I}$ has a *core cover* whenever there exist a series of finite interpretations $\mathcal{I}_1 \subseteq \mathcal{I}_2 \subseteq \ldots$ with $\mathcal{I} = \bigcup_{i>0} \mathcal{I}_i$, such that for all $\mathcal{I}_i$, each homomorphism $h : \mathcal{I}_i \to \mathcal{I}$ is an embedding. Showing there exists a core cover is enough to show our structure is a core.

**Definition 4.2.** Set $can_n(\mathcal{T}, \mathcal{A})$ for each positive natural number $n$ as a *finite approximation* of $can(\mathcal{T}, \mathcal{A})$, as follows:

1. $\Delta^{can_n(\mathcal{T},\mathcal{A})} := \{x \in \Delta^{can(\mathcal{T},\mathcal{A})} \mid |x| \leq n\}$;
2. $a^{can_n(\mathcal{T},\mathcal{A})} := a^{can(\mathcal{T},\mathcal{A})}$, for all $a \in N_I$;
3. $C^{can_n(\mathcal{T},\mathcal{A})} := \{x \in C^{can(\mathcal{T},\mathcal{A})} \mid |x| \leq n\}$, for all $C \in N_C$;
4. $r^{can_n(\mathcal{T},\mathcal{A})} := \{(x, y) \in r^{can(\mathcal{T},\mathcal{A})} \mid |x| \leq n, |y| \leq n\}$, for all $r \in N_R^+$.

Note that $can_n(\mathcal{T}, \mathcal{A})$ is not per se a model of $(\mathcal{T}, \mathcal{A})$. However, if the $n$th approximation is a model, then $can_n(\mathcal{T}, \mathcal{A}) = can(\mathcal{T}, \mathcal{A})$.

**Example 4.3.** Suppose $\mathcal{A} = \{A(a)\}$ and $\mathcal{T} = \{A \sqsubseteq \exists r.A\}$. Let $t = (\{A\}, \{r\}, \{A\})$ be a 2-type. First, notice that $t \in child_{\mathcal{T}}(t)$. Then the $n$th approximation is a chain of length $n$: $\Delta^{can_n(\mathcal{T},\mathcal{A})} = \{at^i \mid i < n\}$, $A^{can_n(\mathcal{T},\mathcal{A})} = \{at^i \mid i < n\}$ and $r^{can_n(\mathcal{T},\mathcal{A})} := \{(at^i, at^{i+1}) \mid i < n-1\}$, where $t^i$ denotes the concatenation of $i$ times $t$. ♡

**Theorem 4.4.** *For each normalised Horn-$\mathcal{ALCHIQ}$ knowledge base $(\mathcal{T}, \mathcal{A})$, $can(\mathcal{T}, \mathcal{A})$ is a core.*

**Proof.** From [29], Theorem 16, we learn that each interpretation that has a core cover is a core. Thus, it suffices to show that $\mathcal{I}_i = can_i(\mathcal{T}, \mathcal{A})$ is a core cover for

$$\mathcal{I} = \bigcup_{i>0} can_i(\mathcal{T}, \mathcal{A}) = can(\mathcal{T}, \mathcal{A}).$$

It is immediate that for each $i$, the identity mapping $h_i : \Delta^{\mathcal{I}_i} \to \Delta^{can(\mathcal{T},\mathcal{A})}$ given by $x \mapsto x$ is an embedding.

Consider a fixed $i$. By induction on $|x|$ it is shown that for all homomorphisms $h : \Delta^{\mathcal{I}_i} \to \Delta^{can(\mathcal{T},\mathcal{A})}$, $h(x) = h_i(x) = x$ holds.

- $|x| = 1$. As by definition, $\{x \in \Delta^{\mathcal{I}_i} \mid |x| = 1\} = \Delta^{\mathcal{I}_i} \cap N_C$, it follows directly that for each homomorphism $h$, $h(x) = x$.
- $|x| \leq n + 1 \leq i$. Let us consider some $ak_1 \ldots k_n \in \Delta^{\mathcal{I}_i}$. By induction hypothesis, we know that for all homomorphisms $h$, $h(ak_1 \ldots k_{n-1}) = ak_1 \ldots k_{n-1}$. As $(ak_1 \ldots k_{n-1}, ak_1 \ldots k_n) \in r^{\mathcal{I}_i}$ for all roles $r$ in $k_n$, we also have to ensure $(ak_1 \ldots k_{n-1}, h(ak_1 \ldots k_n)) \in r^{can(\mathcal{T},\mathcal{A})}$, which means that we can already restrict $h(ak_1 \ldots k_n) \in \{ak_1 \ldots k_{n-2}\} \cup \{ak_1 \ldots k_{n-1}k'_n \mid k'_n \in child_{\mathcal{T}}(k_{n-1})\}$ for each possible homomorphism $h$. By (R4) of Definition 3.3, we find that $h(ak_1 \ldots k_n) \neq ak_1 \ldots k_{n-2}$. Similarly, (R3) eliminates $h(ak_1 \ldots k_n) = ak_1 \ldots k'_n$ for all $k'_n \neq k_n$. So, the only option is $h(ak_1 \ldots k_n) = ak_1 \ldots k_n$, which concludes the induction step.

Thus, $h_i$ is the only homomorphism $h : \mathcal{I}_i \to can(\mathcal{T}, \mathcal{A})$ and the identity mapping is trivially an embedding, which concludes the proof. ♣

The following corollary can be shown with a similar proof.

**Corollary 4.5.** *Every finite approximation $can_n(\mathcal{T}, \mathcal{A})$ of the austere canonical model is a core.*

Next to being a core, the austere canonical model is also the unique core universal model, up to isomorphism, for each normalised Horn-$\mathcal{ALCHIQ}$ knowledge base. This is proven in the following theorem.

**Theorem 4.6.** *Each consistent normalised Horn-$\mathcal{ALCHIQ}$ knowledge base $(\mathcal{T}, \mathcal{A})$ has a unique (up to isomorphism) universal core model, namely the austere canonical model $can(\mathcal{T}, \mathcal{A})$.*

Before we get into the actual proof, note that the definition of a core was originally solely intended for finite structures. There are multiple ways to define a core for infinite structures that all coincide for finite cases [30,31]. One of these approaches, requiring that each endomorphism is an embedding, is the one used up till now in this section. A stronger considered version is to also require each endomorphism to be surjective, that is, requiring that each endomorphism is an isomorphism. To distinguish this case from the core definitions discussed before, we say such an interpretation is a *strong core*. According to Bauslaugh [30], these stronger cores have the nicest behaviour in the infinite case. However, they might not exist: an infinite chain of blank nodes can be a core and at the same time not a strong core.

In the following proof, we prove that the austere canonical model is also a strong core. Some properties of this stronger definition then help to show that there exists indeed a unique universal core model, up to isomorphism. In line with the definition above, we say an interpretation $\mathcal{I}$ has a *strong core cover* if there exist finite interpretations $\mathcal{I}_1 \subseteq \mathcal{I}_2 \subseteq \ldots$ with $\mathcal{I} = \bigcup_{i>0} \mathcal{I}_i$, such that for all $\mathcal{I}_i$, each homomorphism $h : \mathcal{I}_i \to \mathcal{I}$ is an isomorphism.

**Proof.** First, note that the proof of Theorem 16 in [29] can easily be extended to also hold in this case: if an instance has a strong core cover then it is a strong core.

From the proof of Theorem 4.4, it is clear that $\mathcal{I}_i = can_i(\mathcal{T}, \mathcal{A})$ is a strong core cover for $can(\mathcal{T}, \mathcal{A})$, and thus that $can(\mathcal{T}, \mathcal{A})$ is a strong core. Now suppose there exists another universal core model $\mathcal{J}$ of the knowledge base $(\mathcal{T}, \mathcal{A})$. Because both models are universal, there exists homomorphisms in both directions. Since each endomorphism of $can(\mathcal{T}, \mathcal{A})$ must be bijective and strong, and each one of $\mathcal{J}$ injective and strong, it is straightforward to conclude that the homomorphism mapping $\mathcal{J}$ to $can(\mathcal{T}, \mathcal{A})$ must be bijective and strong too: $\mathcal{J}$ and $can(\mathcal{T}, \mathcal{A})$ are indeed isomorphic, as required. ♣

**Corollary 4.7.** *Each consistent normalised Horn-$\mathcal{ALCHIQ}$ knowledge base $(\mathcal{T}, \mathcal{A})$ has a unique universal strong core model, namely $can(\mathcal{T}, \mathcal{A})$.*

### 4.2. Core chase

The core chase is a method that is complete for finding the unique (up to isomorphism) *finite* universal core model, whenever it exists [26]. We specifically note that the core chase never produces an *infinite* structure, that is, it is only relevant to compare the austere canonical model to the result of the core chase in case the core chase terminates and produces a finite universal core model. In fact, we will see that in that case, the results coincide. Following this same paper, we will now briefly introduce the core chase. Recall we may view any set of atoms $\mathcal{A}$ as an interpretation with domain $\Delta^{\mathcal{A}}$. As chase procedures are often seen as series of sets of atoms, this assumption makes the following definitions more in line with the literature on chase procedures.

The core chase is constructed by alternating the application of two operations: firing all not-yet satisfied axioms and taking the core of the resulting structure. This procedure is repeated until it terminates, producing a series of sets of atoms where the last set is defined to be the result of the core chase. If the series does not terminate, the core chase is undefined.

We start by formalising the first operation. For this, we consider a function that fires the right-hand side of axioms: the function $f_x$, defined for each $x \in N_I \cup N_B$. This function translates a concept into a set of atoms, and is inductively defined in the following way: $f_x(\top) := \emptyset$, $f_x(A) := \{A(x)\}$, $f_x(\exists(r_0 \sqcap \ldots \sqcap r_n).C) := \{r_0(x,y), r_0^-(y,x), \ldots, r_n(x,y), r_n^-(y,x)\} \cup f_y(C)$ for some fresh variable $y$ and $f_x(C \sqcap C') := f_x(C) \cup f_x(C')$.

To determine which axioms we want to fire, we define a set of matches $m(\mathcal{T}, \mathcal{A})$ containing pairs of one or two nodes in $N_I \cup N_B$ and a Horn-$\mathcal{ALCHIQ}$ axiom. We say $(c, C \sqsubseteq D) \in m(\mathcal{T}, \mathcal{A})$ iff there exists an axiom $C \sqsubseteq D \in \mathcal{T}$ and a homomorphism $h$ from $f_x(C)$ to $\mathcal{A}$, such that $h(x) = c$ and there is no homomorphism from $f_x(D)$ to $\mathcal{A}$ such that $h(x) = c$. Similarly, we say $((c,d), r \sqsubseteq r') \in m(\mathcal{T}, \mathcal{A})$ iff there exists an axiom $r \sqsubseteq r' \in \mathcal{T}$ such that $r(c,d) \in \mathcal{A}$ and $r'(c,d) \notin \mathcal{A}$. No other elements are contained in $m(\mathcal{T}, \mathcal{A})$.

Combining the above, we can fully define the first operation: for an normalised Horn-$\mathcal{ALCHIQ}$ TBox $\mathcal{T}$ and ABoxes $\mathcal{A}, \mathcal{A}'$, we let $\mathcal{A} \xrightarrow{\mathcal{T}} \mathcal{A}'$ iff

$$\mathcal{A}' = \mathcal{A} \cup \bigcup_{(c, C \sqsubseteq D) \in m(\mathcal{T}, \mathcal{A})} f_c(D) \cup \bigcup_{((c,d), r \sqsubseteq r') \in m(\mathcal{T}, \mathcal{A})} r'(c,d).$$

If there exists an axiom of the form $A \sqsubseteq \geq_1 r.B$ in $\mathcal{T}$, such that $\{A(x), r(x,y), B(y), r(x,z), B(z)\}$ would be contained in $\mathcal{A}'$ for some $\{x,y,z\} \subseteq N_I \cup N_B$, a simple substitution of all $y$'s by $z$'s is performed to get the final $\mathcal{A}'$, assuming $y \notin N_I$.

Furthermore, for any binary relation $\rightarrow$, we use the notation $\mathcal{A}\ (\rightarrow)^{\omega}\ \mathcal{A}'$ to denote that there exists a natural number $n$ such that $\mathcal{A}(\rightarrow)^n\mathcal{A}'$ and for all $\mathcal{A}''$ such that $\mathcal{A}' \rightarrow \mathcal{A}''$ we find $\mathcal{A}' = \mathcal{A}''$. With $\circ$ we denote the concatenation of two binary relations, that is, $\mathcal{A} \rightarrow \circ \rightarrow' \mathcal{B}$ iff there exists $\mathcal{A}'$ such that $\mathcal{A} \rightarrow \mathcal{A}'$ and $\mathcal{A}' \rightarrow' \mathcal{B}$, for each pair of binary relations $\rightarrow$ and $\rightarrow'$. Recall we write $\mathcal{A} \xrightarrow{core} \mathcal{B}$ in case $\mathcal{B}$ is the core of $\mathcal{A}$.

**Definition 4.8** ([26]). Given a normalised Horn-$\mathcal{ALCHIQ}$ knowledge base $(\mathcal{T}, \mathcal{A})$, the *core chase* is the unique, up to isomorphism, set of atoms $\mathcal{B}$ such that $\mathcal{A}\ (\xrightarrow{\mathcal{T}} \circ \xrightarrow{core})^{\omega}\ \mathcal{B}$.

Clearly, this series is not guaranteed to be finite, see for instance the following example.

**Example 4.9.** Suppose $\mathcal{A} = \{A(a)\}$ and $\mathcal{T} = \{A \sqsubseteq \exists r.A\}$. Then

$$\mathcal{A}\ (\xrightarrow{\mathcal{T}} \circ \xrightarrow{core})^n\ \{A(a), r(a,a_1), A(a_1), r(a_1,a_2), A(a_2), \ldots, r(a_{n-1},a_n), A(a_n)\}$$

for each natural number $n$. ♡

As mentioned before, in non-terminating cases, the core chase does not produce a result. Simply taking the union of the sets in the core chase construction is in any case not a good idea. More on this topic is discussed in [29], combined with a proposal on how to generalise the core chase construction to an infinite setting: the stable chase. A drawback of this approach is that the result of the stable chase in general misses the universality property. It is unclear whether the stable chase for description logics might result in an non-universal model too. In case the result of the stable chase for normalised Horn-$\mathcal{ALCHIQ}$ is indeed universal, this result coincides with the austere canonical model, up to isomorphism, which can be shown in a similar fashion as Theorem 4.6.

To conclude, drawbacks of the core chase construction are that it is unclear whether the core chase terminates and produces a result. Furthermore, the process of performing the core chase construction itself is expensive, as taking the core of a structure is expensive and happens regularly. The good news is that for normalised Horn-$\mathcal{ALCHIQ}$, the austere canonical model coincides with the result of the core chase, if existent. This means the above troubles are resolved: the austere canonical model is always defined and is constructed without having to take cores of structures.

$$
\begin{aligned}
c^{\mathcal{I},S} &= \{c^{\mathcal{I}}\} \\
s^{\mathcal{I},S} &= \{e \in \Delta^{\mathcal{I}} \mid s(e) \in S\} \\
(\neg s)^{\mathcal{I},S} &= \{e \in \Delta^{\mathcal{I}} \mid s(e) \notin S\} \\
A^{\mathcal{I},S} &= A^{\mathcal{I}} \\
(\varphi \vee \varphi')^{\mathcal{I},S} &= (\varphi)^{\mathcal{I},S} \cup (\varphi')^{\mathcal{I},S} \\
(\varphi \wedge \varphi')^{\mathcal{I},S} &= (\varphi)^{\mathcal{I},S} \cap (\varphi')^{\mathcal{I},S} \\
(\exists(\textstyle\bigsqcap R).\varphi)^{\mathcal{I},S} &= \{e \in \Delta^{\mathcal{I}} \mid \exists e' \in \Delta^{\mathcal{I}} \forall r \in R : (e,e') \in r^{\mathcal{I}} \wedge e' \in \varphi^{\mathcal{I},S}\} \\
(\exists E.\varphi)^{\mathcal{I},S} &= \{e \in \Delta^{\mathcal{I}} \mid \exists e' \in \Delta^{\mathcal{I}} : (e,e') \in E^{\mathcal{I}} \wedge e' \in \varphi^{\mathcal{I},S}\} \\
(c \wedge \mathsf{eq}(E,E'))^{\mathcal{I},S} &= \begin{cases} \{c^{\mathcal{I}}\} & \text{if } \{e \in \Delta^{\mathcal{I}} \mid (c,e) \in E^{\mathcal{I}}\} = \{e \in \Delta^{\mathcal{I}} \mid (c,e) \in E'^{\mathcal{I}}\} \\ \emptyset & \text{otherwise.} \end{cases} \\
(c \wedge \mathsf{disj}(E,E'))^{\mathcal{I},S} &= \begin{cases} \{c^{\mathcal{I}}\} & \text{if } \{e \in \Delta^{\mathcal{I}} \mid (c,e) \in E^{\mathcal{I}}\} \cap \{e \in \Delta^{\mathcal{I}} \mid (c,e) \in E'^{\mathcal{I}}\} = \emptyset \\ \emptyset & \text{otherwise.} \end{cases}
\end{aligned}
$$

**Fig. 4.** Evaluating shape expressions.

**Theorem 4.10.** *Given a normalised Horn-$\mathcal{ALCHIQ}$ knowledge base $(\mathcal{T},\mathcal{A})$ such that $can(\mathcal{T},\mathcal{A})$ is finite, and let $\mathcal{B}$ be the unique, up to isomorphism, structure such that $\mathcal{A}(\xrightarrow{\mathcal{T}} \circ \xrightarrow{core})^{\omega}\mathcal{B}$, then*

$$\mathcal{B} \cong can(\mathcal{T},\mathcal{A}).$$

**Proof.** We note that $\mathcal{B}$ is a core by definition, and the same holds for $can(\mathcal{T},\mathcal{A})$ by Theorem 4.4. Since both are cores, it suffices to show that there exists homomorphisms in both directions.

Note that $can(\mathcal{T},\mathcal{A})$ is a model by Proposition 3.11. As $\mathcal{B}$ is a universal model [26], it follows that there exists a homomorphism from $\mathcal{B}$ into each model, including $can(\mathcal{T},\mathcal{A})$. By Proposition 3.11 we also have that $can(\mathcal{T},\mathcal{A})$ is universal, which means also the other required homomorphism must exist. ♣

## 5. Recursive SHACL

In the third section, we described non-recursive SHACL validation in presence of ontologies. We will continue the rest of the paper considering a fragment of SHACL that allows recursion: *stratified SHACL*. There is no unique way to extend the simple semantics we discussed before to the recursive case, and some alternatives have been explored in the literature, like the stable model semantics [21], the supported model semantics [25] and the well-founded semantics [22]. The semantics we will discuss here is the *least fixed point* semantics. We note it coincides with the stable model semantics and the well-founded semantics on each set of stratified constraints.

As discussed, recursive SHACL allows shape names in the definition of shape constraints. To this end, let $N_S$ denote a countably infinite set of *shape names* that is disjoint from $N_C \cup N_R \cup N_I \cup N_B$. A *shape constraint* is an expression of the form $s \leftarrow \varphi$, where $s \in N_S$ and $\varphi$ a shape expression defined like in the preliminaries, with the addition of allowing referencing shape names $s$ within $\varphi$. We call $s$ the *head* of a constraint $s \leftarrow \varphi$. Note that we allow $s$ to appear as the head of multiple constraints. In the rest of this article, we impose some additional restrictions and consider only a fragment of recursive SHACL. We leave the generalisation of our results to full recursive SHACL for future work.

That is, let $\varphi$ be defined in the following way

$$\varphi ::= c \mid s \mid \neg s \mid A \mid \varphi \vee \varphi \mid \varphi \wedge \varphi \mid \exists(\textstyle\bigsqcap R).\varphi \mid \exists E.\varphi \mid c \wedge \mathsf{eq}(E,E') \mid c \wedge \mathsf{disj}(E,E'),$$

where $c \in N_I$, $s \in N_S$, $A \in \overline{N_C}$, $R \subseteq \overline{N_R}$, and $E$ and $E'$ regular expressions over the alphabet $\overline{N_R}$.

First, note that our versions of path equality and disjointness, $c \wedge \mathsf{eq}(E,E')$ and $c \wedge \mathsf{disj}(E,E')$, distinguish themselves from the version in the preliminaries and the SHACL standard in two ways. First, the standard requires $E' \in \overline{N_R}$. We generalise this and allow $E'$ to be any regular expression. Secondly, for technical reasons, we add a 'guard' in the form of an individual $c \in N_I$. Furthermore, this fragment lacks counting over regular path expressions and closure constraints. We note that our results do not simply extend to the setting without these restrictions; especially counting over regular paths is a feature that requires a lot of attention, and that no results for the unrestricted setting are known in the literature. We refer the reader to Section 10 for a more extensive discussion. Lastly, we increase the expressivity of SHACL slightly by adding conjunction of roles in the form $\exists R.\varphi$, for $R$ a conjunction of roles in $\overline{N_R}$, which we use to reduce Horn-$\mathcal{ALCHIQ}$-reasoning combined with SHACL to plain SHACL reasoning.

The semantics of recursive SHACL is defined in terms of a *shape assignment.* A shape assignment is a set of shape atoms $S$, such that the rules specified in Fig. 4 are satisfied, taken into account that $s \leftarrow \varphi \in C$ implies that $(\varphi)^{\mathcal{I},S} \subseteq s^{\mathcal{I},S}$. As mentioned before, there is no agreement on a unique semantics for recursive SHACL in the literature. Nevertheless, what all semantics have in common is that the shape assignment $S$, a set of *shape atoms*, of the form $s(c)$, for $s \in N_S$ and $c \in N_I$, satisfies this same set of rules.

In the recursive setting, the form of targets is changed: we now consider a set of shape atoms $s(c)$. We will discuss this further in Section 10. As we are considering shape atoms, we also need constraints providing meaning to the shape names. Thus, the notion of targets is replaced by *shape graphs*. A shape graph is a pair $(\mathcal{C}, \mathcal{G})$, where $\mathcal{C}$ is a set of shape constraints and $\mathcal{G}$ is a set of shape atoms.

### 5.1. Stratified SHACL

In this article, we are interested in shape assignments where each shape atom has a proper justification. That is, we do not wish to consider any set of shape atoms $S$, but put some constraints on the type of shape assignments we accept. One option would be to resort to the full stable model semantics, but only if we want to give up on polynomial time data complexity. Thus, we focus on a more straight-forward semantics, based on *stratified* sets of constraints. This is a fragment of recursive SHACL which supports negation to a certain extend, but restricts the full combination of recursion and negation. To this end and following the logic programming literature [20], a partition of constraints (stratification) is defined such that a justified shape assignment can be constructed by processing each partition individually.

**Definition 5.1.** We say a shape name $s$ *occurs negatively* in a shape constraint $s' \leftarrow \varphi$ if $\neg s$ occurs in $\varphi$. We say a shape name $s$ is *defined* in a set $\mathcal{C}$ of constraints if $s \leftarrow \varphi \in \mathcal{C}$ for some $\varphi$.

A set $\mathcal{C}$ of constraints is *stratified* if it can be partitioned into sets $\mathcal{C}_0, \dots, \mathcal{C}_k$, called *strata*, such that, for all $0 \le i \le k$, the following hold.

1. If $i < k$ and $s'$ occurs in $\varphi$ for some $s \leftarrow \varphi \in \mathcal{C}_i$, then $s'$ is not defined in $\mathcal{C}_{i+1} \cup \dots \cup \mathcal{C}_k$.
2. If $s'$ occurs negatively in $\varphi$ for some $s \leftarrow \varphi \in \mathcal{C}_i$, then $s'$ is not defined in $\mathcal{C}_i \cup \dots \cup \mathcal{C}_k$.

A set of constraints is *stratified* if it admits a stratification.

Without loss of generality, we can assume that all constraints with the same head are defined in the same stratum.

A standard way to obtain a stratification is to define a dependency graph $(V, E, E^*)$ with the set of shape names used as the nodes $V$. In this graph, there are two types of edges: marked edges $E^*$ and standard edges $E$, such that $E^* \subseteq E$. We use the standard edges to mark that $s$ occurs in a shape constraint with head $s'$. In that case, we set $(s, s') \in E$. If this is not just any occurrence, but $s$ occurs *negatively* in a shape constraint with head $s'$, then we also set $(s, s') \in E^*$. The lowest stratum is the biggest set of nodes $X$ such that if $s \in X$ and $(s', s) \in E$, then $s' \in X$ and for all $s \in X$, there does not exist any $s' \in V$ such that $(s', s) \in E^*$. To find the next lowest stratum, simply remove all shape names in $X$ from the dependency graph and repeat the above. This may be repeated until all shape names are assigned a stratum.

Given any stratification of a set of constraints, we compute the shape assignment of stratified SHACL with a least fixed point operator over each stratum, with negation as failure to compute the opposite in an earlier stratum. To do so, we first define the notion of an immediate consequence operator $T_{\mathcal{I},\mathcal{C}}$ that, given a shape assignment $S$, adds new shape atoms to satisfy the constraints that are fired by the constraints in $\mathcal{C}$ based on $S$ and $\mathcal{I}$.

**Definition 5.2.** Given a set of constraints $\mathcal{C}$ and an interpretation $\mathcal{I}$ with $N_I \subseteq \Delta^{\mathcal{I}}$, we define an immediate consequence operator $T_{\mathcal{I},\mathcal{C}}$ that maps shape assignments to shape assignments as follows:

$$T_{\mathcal{I},\mathcal{C}}(S) := S \cup \{ s(a) \mid s \leftarrow \varphi \in \mathcal{C}, a \in (\varphi)^{\mathcal{I},S} \}.$$

The following two propositions are a direct consequence of the characterisations from [20] in the context of stratified logic programs. Here, we will again use the definition of the least fixed point starting from a given set. We refer the reader back to Definition 3.6 for the precise definition.

**Proposition 5.3.** *If $\mathcal{C}$ is a constraint set that does not define any shape names that occur negatively in $\mathcal{C}$, then the following hold:*

1. *$T_{\mathcal{I},\mathcal{C}}$ is monotonic, i.e. if $S \subseteq S'$, then $T_{\mathcal{I},\mathcal{C}}(S) \subseteq T_{\mathcal{I},\mathcal{C}}(S')$;*
2. *$T_{\mathcal{I},\mathcal{C}}$ is finitary, i.e. $T_{\mathcal{I},\mathcal{C}}(\bigcup_{n=0}^{\infty} S_n) \subseteq \bigcup_{n=0}^{\infty} T_{\mathcal{I},\mathcal{C}}(S_n)$ for all infinite sequences $S_0 \subseteq S_1 \subseteq \dots$;*
3. *$T_{\mathcal{I},\mathcal{C}}$ is growing, i.e. $T_{\mathcal{I},\mathcal{C}}(S_2) \subseteq T_{\mathcal{I},\mathcal{C}}(S_3)$ for all $S_1, S_2, S_3$ such that $S_1 \subseteq S_2 \subseteq S_3 \subseteq T_{\mathcal{I},\mathcal{C}} \uparrow^{\omega} (S_1)$.*

**Proposition 5.4.** *If $\mathcal{I}$ is an interpretation and $\mathcal{C}_0, \dots, \mathcal{C}_k$ is a stratification of $\mathcal{C}$, then each $T_{\mathcal{I},\mathcal{C}_0}, \dots, T_{\mathcal{I},\mathcal{C}_k}$ is monotone, finitary, and growing. Thus, for any shape assignment $S$ and each $0 \le j \le k$, $T_{\mathcal{I},\mathcal{C}_j} \uparrow^{\omega} (S)$ is a fixpoint of $T_{\mathcal{I},\mathcal{C}_j}$.*

Based on the above, we can now define the computation of the desired shape assignment along a stratification $\mathcal{C}_0, \dots, \mathcal{C}_k$ of $\mathcal{C}$.

**Definition 5.5.** Assume $\mathcal{I}$ is an interpretation, $\mathcal{C}$ is a stratified set of constraints, and let $\mathcal{C}_0, \dots, \mathcal{C}_k$ be a stratification of $\mathcal{C}$. Then let

$$M_0 := T_{\mathcal{I},\mathcal{C}_0} \uparrow^{\omega}(\emptyset)$$

$$M_i := T_{\mathcal{I},\mathcal{C}_i} \uparrow^{\omega}(M_{i-1}) \quad \text{for each } 1 \le i < k.$$

We call $M_k$ be the *perfect assignment* for $\mathcal{C}$ and $\mathcal{I}$, and use the notation $PA(\mathcal{C}, \mathcal{I}) := M_k$. An interpretation $\mathcal{I}$ (resp., ABox $\mathcal{A}$) *validates* a shapes graph $(\mathcal{C}, \mathcal{G})$ if $\mathcal{G} \subseteq PA(\mathcal{C}, \mathcal{I})$ (resp., $\mathcal{G} \subseteq PA(\mathcal{C}, \mathcal{I}_{\mathcal{A}})$).

A well-known result is that the particular stratification chosen does not matter: any stratification will give the same perfect assignment [20]. Given this semantics, it is now straightforward to extend Definition 3.12 to recursive SHACL and shapes graphs.

**Definition 5.6.** Given a normalised Horn-$\mathcal{ALCHIQ}$ TBox $\mathcal{T}$ and ABox $\mathcal{A}$, and a shapes graph $(\mathcal{C}, \mathcal{G})$, we say $(\mathcal{T}, \mathcal{A})$ validates $(\mathcal{C}, \mathcal{G})$ if $can(\mathcal{T}, \mathcal{A})$ validates $(\mathcal{C}, \mathcal{G})$.

### 5.2. Normal form

A nice feature of the SHACL fragment introduced at the start of this section, is that it has a rather succinct normal form with the same expressivity. In the rest of the paper, we will assume that all the given sets of constraints are in normal form. This will be particularly useful in the following two sections, restricting the shapes of constraints we need to consider.

**Definition 5.7.** A SHACL constraint is in *normal form* if it has one of the following forms

| | |
|---|---|
| (NC1) $s \leftarrow c$ | (NC4) $s \leftarrow s' \wedge s''$ |
| (NC2) $s \leftarrow s'$ | (NC5) $s \leftarrow \exists(\bigsqcap R).s'$ |
| (NC3) $s \leftarrow A$ | (NC6) $s \leftarrow \neg s'$, |

where $c \in N_I$, $R \subseteq \overline{N_R}$, $\{s, s', s''\} \subseteq N_S$ and $A \in \overline{N_C}$.

We use the term *negative constraints* to refer to constraints of the form (NC6) and *positive constraints* for constraints of the form (NC1)–(NC5). Clearly, each set of positive constraints is automatically stratified. Furthermore, for simplicity, if $R = \{r\}$, we may write $s \leftarrow \exists r.s'$ instead of $s \leftarrow \exists(\bigsqcap R).s'$.

We can indeed rewrite parts of SHACL into this restricted, normalised, form of SHACL, as formalised below.

**Proposition 5.8.** *Each set of constraints $C$ can be translated in time polynomial in $|C|$, into a set of constraints $C'$ in normal form such that for all $\mathcal{G}$ and each normalised Horn-$\mathcal{ALCHIQ}$ knowledge base $(\mathcal{T}, \mathcal{A})$, we have $(\mathcal{T}, \mathcal{A})$ validates $(C, \mathcal{G})$ iff $(\mathcal{T}, \mathcal{A})$ validates $(C', \mathcal{G})$.*

In particular, the above statement holds for $\mathcal{T} = \emptyset$, that is, the standard case of SHACL validation, without ontologies. Furthermore, normalising a set of constraints in the way described below does not influence it being stratified or not.

**Proof.** We extend the results for normal forms of [21] and [32]. That is, we first recursively introduce fresh shapes for sub-expressions that appear in constraints, as in [21].

Next to that, we also allow shape names to appear as heads in multiple constraints, which means that $s \leftarrow \varphi \vee \varphi'$ can be replaced by $s \leftarrow \varphi$ and $s \leftarrow \varphi'$ without affecting validation. Now, what is left to show is that also constraints of the form $s \leftarrow c \wedge \mathsf{eq}(E, E')$, $s \leftarrow c \wedge \mathsf{disj}(E, E')$ and $s \leftarrow \exists E.s'$ can be translated into constraints in normal form in polynomial time. The last case, $s \leftarrow \exists E.s'$, we already covered in [32], but is repeated here for completeness.

- $s \leftarrow \exists E.s'$. Suppose $\mathcal{M} = (Q, \Sigma, q_I, \Delta, q_F)$ is the automaton recognising $E$. Take fresh shape names $s_q$ for each $q \in Q$ and add the following constraints

  $s \leftarrow s_{q_I}$
  $s_q \leftarrow \exists r.s'_q \quad$ if $(q, r, q') \in \Delta$
  $s_{q_F} \leftarrow s'$.

  Here, the idea is to encode the states of the automaton in the set of shape names. However, we do not consider them in the standard initial state towards final state mode, but the other way around. In the end, the idea is that we let every state that is an $s'$ be a final state, and see whether we can work our way back through the automaton to an initial state. Thus, we find that a node is assigned $s_{q_I}$ iff that node has an $E$-path to a node that is an $s'$.
- $s \leftarrow c \wedge \mathsf{eq}(E, E')$. Suppose $\mathcal{M} = (Q, \Sigma, q_I, \Delta, q_F)$ and $\mathcal{M}' = (Q', \Sigma, q'_I, \Delta', q'_F)$ are the automata recognising $E$ and $E'$ respectively, such that $Q \cap Q' = \emptyset$. Take fresh shape names $s_{error}$, $s_{noerror}$ $s_{pos}$, $s_{neg}$, and $s_q$ for each $q \in Q \cup Q'$ and add the following constraints

  | | |
  |---|---|
  | $s_q \leftarrow c$ | if $s_q \in \{s_{q_I}, s_{q'_I}\}$ |
  | $s_{q'} \leftarrow \exists r^-.s_q$ | if $(q, r, q') \in \Delta \cup \Delta'$ |
  | $s_{pos} \leftarrow s_q$ | if $s_q \in \{s_{q_F}, s_{q'_F}\}$ |
  | $s_{neg} \leftarrow \neg s_q$ | if $s_q \in \{s_{q_F}, s_{q'_F}\}$ |
  | $s_{error} \leftarrow s_{pos} \wedge s_{neg}$ | |
  | $s_{error} \leftarrow \exists r.s_{error}$ | if $r \in \Sigma$ |
  | $s_{noerror} \leftarrow \neg s_{error}$ | |
  | $s \leftarrow s_q \wedge s_{noerror}$. | |

  In this case, because of the 'guard' $c$, there is only one node considered to be the initial state of both automata. From there, we assign the states using their encoding in shape names. This time, we consider the states of the automaton from initial state towards, possibly, a final state somewhere. After finishing assigning states, it is decided which nodes are the final state of any of the two automata. The last thing to do is to see whether there exists a node that is the final state of one of the two automata (thus being assigned $s_{pos}$), but not of the other automaton (thus also being assigned $s_{neg}$). The last thing to do is to bring this information back towards $c$, with the constraints $s_{error} \leftarrow \exists r.s_{error}$ for each $r$ appearing in the automata.
- $s \leftarrow c \wedge \mathsf{disj}(E, E')$. Similar solution as $s \leftarrow c \wedge \mathsf{eq}(E, E')$, but define the error shape $s_{error}$ as $s_{error} \leftarrow s_{q_F} \wedge s_{q'_F}$.

Note that the automata can be constructed in polynomial time, which is a standard result in the literature, see for instance [33]. ♣

## 6. Rewriting for positive SHACL

Now we will return to problem of SHACL validation in presence of ontologies. In Section 3, we defined the semantics of SHACL in presence of ontologies as SHACL validation over the core universal model. This was independent of the chosen semantics for SHACL.

As we decided on the least fixed point semantics for SHACL in the previous section, it is now clear what the intended meaning is of a knowledge base validating a shapes graph.

However, we also care about computability. To this end, we do not wish to materialise the full, possibly infinite, core universal model. Instead, we will bring SHACL validation over the core universal model back to SHACL validation over the enriched ABox $\mathcal{A}_\mathcal{T}$, as defined in Definition 3.7, by rewriting the set of constraints. More precisely, given a TBox $\mathcal{T}$ and a set of stratified constraints $\mathcal{C}$, we want to compile $\mathcal{T}$ and $\mathcal{C}$ into a new set $\mathcal{C}_\mathcal{T}$ of stratified constraints so that for every ABox $\mathcal{A}$ consistent with $\mathcal{T}$, and every target $\mathcal{G}$, we have

$$(\mathcal{T}, \mathcal{A}) \text{ validates } (\mathcal{C}, \mathcal{G}) \text{ iff } \mathcal{A}_\mathcal{T} \text{ validates } (\mathcal{C}_\mathcal{T}, \mathcal{G}).$$

This will be achieved by means of an inference procedure that uses a collection of inference rules to capture the possible "propagation" of shape names in the anonymous part of the austere canonical model. In the following, the rewriting procedure for SHACL without any occurrence of negation is discussed, followed by the general case of recursive SHACL.

### 6.1. Rewriting algorithm

The idea of the rewriting is that we can encode all reasoning of axioms in the constraints. The core of our technique is an inference procedure that derives a set of quadruples $(t, P, Q, H)$, where $t$ a 2-type, as introduced at the start of Section 3.1. Intuitively, deriving $(t, P, Q, H)$ tells us that an object that satisfies all concept names in $\pi_1(t)$, that satisfies all expressions (assumptions) in $P$, and does not satisfy every single expression in $Q$, must validate all shape names in $H$. Note that there may be some overlap in the information stored in $t$ and $P$.

**Definition 6.1.** Let $\top$, $c$, $\exists r.s$, be *basic shape expressions*, for $r \in \overline{N_R}$, $s \in N_S$, and $c \in N_I$. Moreover, let $\exists(\bigsqcap R).\bigsqcap N$ be called a *basic concept expression*, for each $R \subseteq \overline{N_R}$ and $N \subseteq N_C$.

Syntactically, $P$ and $Q$ are sets of basic shape and basic concept expressions, whereas the set $H$ may only contain shape names in $N_S$.

**Definition 6.2.** A 2-type $t$ is called *locally consistent* if the following are satisfied.

- If $\mathcal{T} \models A_1 \sqcap \ldots \sqcap A_n \sqsubseteq B$ and $\{A_1, \ldots, A_n\} \subseteq \pi_i(t)$, then $B \in \pi_i(t)$, for $i \in \{1, 3\}$.
- If $\mathcal{T} \models \bigsqcap R \sqsubseteq r'$ and $R \subseteq \pi_2(t)$, then $r' \in \pi_2(t)$.
- If $A \sqsubseteq \forall r.B \in \mathcal{T}$, $A \in \pi_1(t)$ and $r \in \pi_2(t)$, then $B \in \pi_3(t)$.
- If $A \sqsubseteq \forall r.B \in \mathcal{T}$, $A \in \pi_3(t)$ and $r^- \in \pi_2(t)$, then $B \in \pi_1(t)$.

Recall that we are only considering consistent knowledge bases. As discussed in the preliminaries, this means we may ignore all axioms containing '$\bot$' in the universal model construction. On the same note, there is also no need to consider those axioms in the following rewriting procedure.

**Definition 6.3.** Given a normalised Horn-$\mathcal{ALCHIQ}$ TBox $\mathcal{T}$ and $\mathcal{C}$ a set of normalised constraints, let $N_C^\mathcal{T} \subseteq N_C$ be the (finite) set of concepts occurring in $\mathcal{T}$ or $\mathcal{C}$. We let $psat_{\mathcal{C},\mathcal{T}}$ be the smallest set of quadruples $(t, P, Q, H)$ that is closed under the following rules.

1. If $t$ is a locally consistent 2-type, and $Q$ is a set of basic shape expressions such that for all $\exists(\bigsqcap R).(\bigsqcap N) \in Q$, we have $(R, N) \in succ_\mathcal{T}(t)$ and not both $R \subseteq \pi_2(t)$ and $N \subseteq \pi_3(t)$, then $(t, \bar{Q}, Q, \emptyset)$ belongs to $psat_{\mathcal{C},\mathcal{T}}$, where
$$\bar{Q} := \{\top\} \cup \{\exists(\bigsqcap R).(\bigsqcap N) \mid (R, N) \in succ_\mathcal{T}((\pi_1(t), \emptyset, \emptyset)), \exists(\bigsqcap R).(\bigsqcap N) \notin Q\}.$$
2. If $\{(t, P, Q, H), (t, P', Q', H')\} \subseteq psat_{\mathcal{C},\mathcal{T}}$ such that $\exists(\bigsqcap R).(\bigsqcap N) \in Q$ iff $\exists(\bigsqcap R).(\bigsqcap N) \in Q'$, then $(t, P \cup P', Q \cup Q', H \cup H')$ belongs to $psat_{\mathcal{C},\mathcal{T}}$.
3. If $s \leftarrow S \in \mathcal{C}$ for some basic shape expression $S$ and $(t, P, Q, H) \in psat_{\mathcal{C},\mathcal{T}}$, then $(t, P \cup (\{S\} \setminus N_C), Q, H \cup \{s\})$ belongs to $psat_{\mathcal{C},\mathcal{T}}$ if either
   - $S = c$, for $c \in N_I$ and $c \notin Q$; or
   - $S = A$, for $A \in \pi_1(t)$; or
   - $S = \exists r.s'$ and $r \in \pi_2(t)$ or there exists $\exists(\bigsqcap R).(\bigsqcap N) \in P$ such that $r \in R$ and $\exists r.s' \notin Q$.
4. If $s \leftarrow s' \in \mathcal{C}$, $(t, P, Q, H) \in psat_{\mathcal{C},\mathcal{T}}$ and $\{s'\} \subseteq H$, then $(t, P, Q, H \cup \{s\})$ belongs to $psat_{\mathcal{C},\mathcal{T}}$.
5. If $s \leftarrow s_1 \wedge s_2 \in \mathcal{C}$, $(t, P, Q, H) \in psat_{\mathcal{C},\mathcal{T}}$ and $\{s_1, s_2\} \subseteq H$, then $(t, P, Q, H \cup \{s\})$ belongs to $psat_{\mathcal{C},\mathcal{T}}$.
6. If $s \leftarrow \exists(\bigsqcap R).s' \in \mathcal{C}$, $\{(t, P, Q, H), (t', P', Q', H')\} \subseteq psat_{\mathcal{C},\mathcal{T}}$, $R^- \subseteq \pi_2(t')$, $P' \cap N_I = \emptyset$, $s' \in H'$ and $inv(t') \in child_\mathcal{T}(inv(t))$ such that
   - $\exists(\bigsqcap R').(\bigsqcap N') \in P'$ iff $(R', N') \in succ_\mathcal{T}((\pi_1(t'), \emptyset, \emptyset)) \setminus succ_\mathcal{T}(t')$; and
   - $\{s'' \mid \exists r''.s'' \in P'\} \subseteq H$,

   then $(t, P, Q, H \cup \{s\})$ belongs to $psat_{\mathcal{C},\mathcal{T}}$.

In this definition, Rule 1 ensures that for each possible node in a data graph, there is a quadruple describing the 1-type of that node in $\pi_1(t)$, restricted to the concept names occurring in $\mathcal{T}$. Furthermore, the set of neighbours this 1-type needs to have in all models is partitioned into a set of neighbours we observe ($P$), and which will be introduced as children of this 1-type ($Q$). Because of this, we may consider the first three entries of the quadruple as a description of a small part of a model, containing all information about its type and neighbours that might be relevant. In this way, it contains enough information to mimic the least fixed point operator for shape names step by step by Rules 3–6. Note that because we are considering the 2-type from the 'perspective' of the node, that is, 'looking back towards the data', we have to consider their inverses when checking the child-relationship in Rule 6. Lastly, Rule 2 is not necessary, but eases the completeness proof. We note that the same set of rules will be reused for the rewriting of stratified SHACL. Because of this, certain conditions do not have a purpose yet, like $c \notin Q$ in Rule 3, which will always be met in the current positive setting.

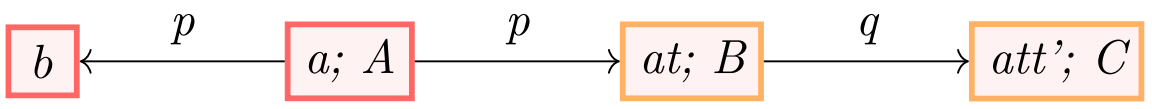

**Fig. 5.** Austere Canonical Model $can(\mathcal{T},\mathcal{A})$, for $\mathcal{A} = \{A(a), p(a,b)\}$ and $\mathcal{T} = \{A \sqsubseteq \exists p.B, B \sqsubseteq \exists q.C\}$.

**Definition 6.4** (Rewriting Procedure)**.** Consider a normalised Horn-$\mathcal{ALCHIQ}$ TBox $\mathcal{T}$, $C$ a set of normalised constraints and $K$ a set of quadruples $(t,P,Q,H)$. Let $C_{\mathcal{T},K}$ be the set of constraints that contains $C$, and moreover, for each $(t,P,Q,H) \in K$ and each $s \in H$, the constraint

$$s \leftarrow \bigwedge_{A\in\pi_1(t)} A \wedge \bigwedge_{A\in N_C^{\mathcal{T}}\setminus\pi_1(t)} \neg A \wedge \bigwedge_{X\in P} X \wedge \bigwedge_{Y\in Q} \neg Y. \qquad (1)$$

Note that only $\pi_1(t)$ and not the full 2-type is taken into account when transforming a triple into a constraint. This suffices as we are considering axioms in normal form. Thus, knowing exactly which concept names are true in a node implies knowing exactly which structures may appear in the anonymous tree structure with this node as its root. The next step is eliminating the superfluous substructures of this anonymous tree, similar to what can be expected in any core universal model construction. This boils down to knowing exactly which of the substructures of depth at most 1 are already present in the enriched ABox $\mathcal{A}_{\mathcal{T}}$. Dividing all relevant substructures in the positive or negative set $P$ and $Q$ happens in Rule 1. As $P$ is subsuming the information stemming from $\pi_2(t)$ and $\pi_3(t)$, it does not have to be repeated in (1).

**Example 6.5.** Suppose $\mathcal{T} = \{A \sqsubseteq \exists p.B, B \sqsubseteq \exists q.C\}$ and consider the following constraints $C$:

$$s \leftarrow \exists p.s \qquad s \leftarrow \exists q.s \qquad s \leftarrow s' \wedge s'' \qquad s' \leftarrow \exists p^-.s' \qquad s' \leftarrow \exists q^-.s' \qquad s' \leftarrow A \qquad s'' \leftarrow C.$$

Let $t = (\{A\},\{p\},\{B\})$ and $t' = (\{B\},\{q\},\{C\})$ and recall the definition of the inverse of a 2-type: $inv(t) = (\{B\},\{p^-\},\{A\})$. Note that both $inv(t)$, $inv(t')$ and $t_A := (\{A\},\emptyset,\emptyset)$ are locally consistent 2-types. Thus, according to the Rule 1 of Definition 6.3, the following quadruples belong to $psat_{C,\mathcal{T}}$:

$$(t_A,\{\top\},\{\exists p.B\},\emptyset), \qquad (inv(t),\{\top\},\{\exists q.C\},\emptyset), \qquad (inv(t'),\{\top\},\emptyset,\emptyset).$$

Given these quadruples and following Rule 3 (of Definition 6.3) multiple times, we can also derive the following quadruples:

$$(t_A,\{\top\},\{\exists p.B\},\{s'\}), \quad (inv(t),\{\top,\exists p^-.s'\},\{\exists q.C\},\{s'\}), \quad (inv(t'),\{\top,\exists q^-.s'\},\emptyset,\{s',s''\}).$$

That is, in the first quadruple, $s'$ is added as a conclusion as $A$ is contained in $\pi_1(t_A)$. For the second quadruple, we add $\exists p^-.s'$ to the set of positive assumptions, thus also describing an environment in which $s'$ can be concluded. To see that the third quadruple is contained in $psat_{C,\mathcal{T}}$, Rule 3 is applied twice: once because $C$ is contained in $\pi_1(inv(t'))$, and once because of the addition $\exists q^-.s'$ to the positive assumptions. The last quadruple may be updated to $(inv(t'),\{\top,\exists q^-.s'\},\emptyset,\{s,s',s''\})$ because of $s \leftarrow s' \wedge s''$. Now Rule 6 can be used based on $s \leftarrow \exists q.s$, $(inv(t),\{\top,\exists p^-.s'\},\{\exists q.C\},\{s'\})$ and the quadruple $(inv(t'),\{\top,\exists q^-.s'\},\emptyset,\{s,s',s''\})$ to infer that $(inv(t),\{\top,\exists p^-.s'\},\{\exists q.C\},\{s,s'\})$ must belong to $psat_{C,\mathcal{T}}$ too. That is, we compare the environments described by these two quadruples: the main idea is that we want to derive $s$ in the first quadruple because of a $q$ connecting the first environment to the second one (reflected by the check $\{q^-\} \subseteq \pi_2(inv(t'))$), where $s$ will be derived ($s \in H'$). The existence of this second environment must be supported by its type being a child of the first type ($inv(inv(t')) \in child_{\mathcal{T}}(inv(inv(t)))$) and thus cannot have an individual name in the set of positive assumptions ($P' \cap N_I = \emptyset$). Lastly, every assumption of the form $\exists r.s$ assumes, in case of an anonymous node, that its parent must be an $s$, and in case of a data node, that there exists an $r$-neighbour that is an $s$. As the second environment is assumed to describe an anonymous node, with as 'parent environment' the first quadruple, we have to check whether $s' \in H$. Similarly, positive assumptions of the form $\exists(\bigsqcap R').(\bigsqcap N')$ must be evaluated 'upwards towards the parent environment' as well, that is $R' \subseteq \pi_2(t')$ and $N' \subseteq \pi_3(t')$. Note that this exactly corresponds to $(R',N') \in succ_{\mathcal{T}}((\pi_1(t'),\emptyset,\emptyset)) \setminus succ_{\mathcal{T}}(t')$ for all $\exists(\bigsqcap R').(\bigsqcap N')$ that may appear in $P'$. Using this derived quadruple and again Rule 6, but now together with the quadruple $(t_A,\{\top\},\{\exists p.B\},\{s'\})$ and constraint $s \leftarrow \exists p.s$, we also find $(t_A,\{\top\},\{\exists p.B\},\{s,s'\}) \in psat_{C,\mathcal{T}}$. Thus, the following constraint is contained in $C_{\mathcal{T},\emptyset}$:

$$s \leftarrow A \wedge \neg B \wedge \neg C \wedge \top \wedge \neg\exists p.B.$$

Now let $\mathcal{A} = \{A(a), p(a,b)\}$. The target we wish to validate is $\mathcal{G} = \{s(a)\}$. It is now simple to check that this target, $s(a)$, is validated by considering the above constraint on $\mathcal{A}_{\mathcal{T}}$, which coincides with $\mathcal{A}$ in this example.

The other approach to test validation is to build $can(\mathcal{T},\mathcal{A})$, graphically depicted in Fig. 5, and then check validation of $(C,\mathcal{G})$. In this figure, the same notation is used as described in Example 3.2. It is straightforward to check that $can(\mathcal{T},\mathcal{A})$ indeed validates $(C,\mathcal{G})$, as required. ♡

We note that, even in the best-case scenario, Definition 6.3 creates an amount of quadruples exponential in size of $\mathcal{T}$ and $C$. This expressive power is used to describe all possible relevant settings that may appear at the border between the data and anonymous part of the austere canonical model. However, what is considered relevant, is only dependent on the TBox $\mathcal{T}$; concept names that do not appear in any axiom cannot be introduced or introduce anything new and may thus be ignored. This means that the exponential blow-up observed in this definition does not influence the data complexity of determining SHACL validation in presence of ontologies. Moreover, as the described SHACL constraint rewriting happens completely independent of the data, the same rewritten shape constraint sets may be reused for different data graphs.

### 6.2. Completeness and correctness

In the rest of this section, we will show completeness and correctness of the presented rewriting technique to translate (positive) SHACL validation in presence of ontologies to plain SHACL validation over the enriched ABox $\mathcal{A}_{\mathcal{T}}$. After that we aim further: in the next subsection, we discuss how to handle stratified negation in the rewriting, followed by a section on how to achieve a pure rewriting, that is, to plain SHACL validation over the original ABox $\mathcal{A}$.

For a quadruple $\rho = (t, P, Q, H)$ a shape name $s \in H$ and a fresh shape name $s_\rho$, we define the constraint $s_\rho$ as follows:

$$s_\rho \leftarrow \bigwedge_{A\in\pi_1(t)} A \wedge \bigwedge_{A\in N_C^{\mathcal{T}}\setminus\pi_1(t)} \neg A \wedge \bigwedge_{X\in P} X \wedge \bigwedge_{Y\in Q} \neg Y.$$

Note that this definition is very close to Definition 6.4. However, it has a different purpose. In the following proofs, we will focus on validating $s_\rho$ instead of $s$. This makes the connection to a certain quadruple precise. And clearly, if $s_\rho(c)$ is contained in some perfect assignment, also $s(c)$ is contained in that same assignment. We will use the notation $C \cup s_\rho$ to denote that the constraint defining $s_\rho$ as above is added to the constraints in $C$.

Furthermore, we note that quadruples are designed to model the direct environment of a node that is on the 'border' of the model under construction. More precisely, they model the environment of $c$ in $can_{|c|}(\mathcal{T}, \mathcal{A})$ (sometimes shortened to $\mathcal{I}_{|c|}$,) the approximation of the austere canonical model, as defined in Definition 4.2. In this interpretation, $c$ is missing its successors. For this reason, we will evaluate $s_\rho(c)$ specifically on $can_{|c|}(\mathcal{T}, \mathcal{A})$. When proving this in general, it definitely also holds for all $c$ such that $|c| = 1$, or put differently, all $c \in N_I$.

To clarify which shape atoms depend on others, we consider the immediate consequence operator defined in Definition 5.2. Where 'depending' is intended in the sense that $s(c)$ depends on $s'(c)$ and $s''(c)$ in case of the constraint $s \leftarrow s' \wedge s''$. Recall that the perfect assignment for positive constraints consists simply of the least fixed point of this operator. With all this machinery, we can demonstrate how the rules in Definition 6.3 can mimic all these types of dependencies. In this way, we show first the completeness, and then, in Proposition 6.8, the correctness of the rewriting for positive constraints.

**Proposition 6.6.** *Given a normalised Horn-$\mathcal{ALCHIQ}$ TBox $\mathcal{T}$ and $C$ a set of positive, normalised constraints, for every target $\mathcal{G}$ and every ABox $\mathcal{A}$ that is consistent with $\mathcal{T}$, we have that if $(\mathcal{T}, \mathcal{A})$ validates $(C, \mathcal{G})$, then $can_0(\mathcal{T}, \mathcal{A})$ validates $(C_{\mathcal{T}}, \mathcal{G})$.*

**Proof.** It suffices to prove the following claim, which will be shown by induction on $n$.

**Claim 6.7.** *Let $\mathcal{I}_i = can_i(\mathcal{T}, \mathcal{A})$ and $S_n = T_{can(\mathcal{T},\mathcal{A}),C} \uparrow^n (\emptyset)$. If $s(c) \in S_{n+1}$, there exists a quadruple $\rho = (t, P, Q, H) \in psat_{C,\mathcal{T}}$ such that $s \in H$, $s_\rho(c) \in T_{\mathcal{I}_{|c|},C_{\mathcal{T}}\cup s_\rho}(S_n)$ and furthermore:*

1. *if $c \in N_{(\mathcal{T},\mathcal{A})} \setminus N_I$, then*
   (a) *$t = inv(tail(c))$;*
   (b) *$P \supseteq \{\exists(\bigsqcap R).(\bigsqcap N) \mid (R, N) \in succ_{\mathcal{T}}((\pi_1(t), \emptyset, \emptyset)) \setminus succ_{\mathcal{T}}(t)\}$ and $P \cap N_I = \emptyset$;*
   (c) *$Q \supseteq \{\exists(\bigsqcap R).(\bigsqcap N) \mid (R, N) \in succ_{\mathcal{T}}((\pi_1(t), \emptyset, \emptyset))\}$;*
2. *for all other $c \in N_I$, we have*
   (a) *$t = (\{A \in N_C \mid A(c) \in \mathcal{A}_{\mathcal{T}}\}, \emptyset, \emptyset)$;*
   (b) *$P \supseteq \{\exists(\bigsqcap R).(\bigsqcap N) \mid (R, N) \in succ_{\mathcal{T}}(t), c \in (\exists(\bigsqcap R).(\bigsqcap N))^{\mathcal{A}_{\mathcal{T}}}\}$ and $(P \cap N_I) \setminus \{c\} = \emptyset$;*
   (c) *$Q \supseteq \{\exists(\bigsqcap R).(\bigsqcap N) \mid (R, N) \in succ_{\mathcal{T}}(t), c \notin (\exists(\bigsqcap R).(\bigsqcap N))^{\mathcal{A}_{\mathcal{T}}}\}$;*

For $(n = 0)$, note that either $s \leftarrow c \in C$ or $s \leftarrow A \in C$. In the first case, $\rho = (t, P \cup \{c\}, Q, \{s\})$ can be constructed by performing Rule 1, followed by Rule 3, where $t = (\{A \in N_C \mid A(c) \in \mathcal{A}_{\mathcal{T}}\}, \emptyset, \emptyset)$, and $P$ and $Q$ are the smallest sets that satisfy the criteria. Removing $\{c\}$ from the second entry in the quadruple, and possibly using $t = inv(tail(c))$ instead of $t = (\{A \in N_C \mid A(c) \in \mathcal{A}_{\mathcal{T}}\}, \emptyset, \emptyset)$ in case $c \notin N_I$, suffices for the second case. It is then easy to check that indeed $s_\rho(c) \in T_{\mathcal{I}_0,C_{\mathcal{T}}\cup s_\rho}(\emptyset)$.

Now suppose $s(c) \in S_{n+1} \setminus S_n$. In this induction step, we will only focus on the case $s(c)$ got derived from $s \leftarrow \exists r.s' \in C$ and $s'(d) \in S_n$, for some $r$-neighbour $d$ of $c$, as the other derivation options are reflected in a straightforward way in the rules of Definition 6.3. We distinguish two cases: $|d| \leq |c|$, which we call 'upwards', and $|d| > |c|$, which we call 'downwards'.

For the upwards case, if $|d| \leq |c|$ then either there exists a 2-type $t$ such that $dt = c$ and $r^- \in \pi_2(t)$, or $\{c, d\} \subseteq N_I$.

- $dt = c$. By Rule 1, we find $(inv(t), P, Q, \emptyset) \in psat_{C,\mathcal{T}}$, such that 1a, 1b and 1c of the claim are satisfied. With Rule 3, recalling that we assumed $s \leftarrow \exists r.s' \in C$, we can update this quadruple to find $\rho = (inv(t), P \cup \{\exists r.s'\}, Q, \{s\}) \in psat_{C,\mathcal{T}}$. As $s'(d) \in S_n$, it is straightforward to see that $s_\rho(c) \in T_{\mathcal{I}_{|c|},C_{\mathcal{T}}\cup s_\rho}(S_n)$, which concludes this case.

- $\{c,d\} \subseteq N_I$. Again, by Rule 1, we find $(t,P,Q,\emptyset) \in psat_{C,\mathcal{T}}$, where $t = (\{A \in N_C \mid A(c) \in \mathcal{I}_0\}, \emptyset, \emptyset)$ and such that 1a, 1b and 1c of the claim are satisfied. Applying Rule 3, means that also $\rho = (t, P \cup \{\exists r.s'\}, Q, \{s\}) \in psat_{C,\mathcal{T}}$. With similar reasoning as in the previous case, it is clear that indeed $s_\rho(c) \in T_{\mathcal{I}_0, C_\mathcal{T} \cup s_\rho}(S_n)$.

In the second case, downwards, we have $|d| > |c|$, that is, $ct' = d$, for some 2-type $t'$. In the following argument, we assume $c \in N_{(\mathcal{T},\mathcal{A})} \setminus N_I$, but a similar argument works for $c \in N_I$. Because we assumed that $s'(d) \in S_n$, we can apply the induction hypothesis to find $\rho' = (inv(t'), P', Q', H') \in psat_{C,\mathcal{T}}$ such that (i) $r^- \in \pi_2(inv(t'))$, (ii) $P' \cap N_I = \emptyset$, (iii) $s' \in H'$, (iv) $\exists(\bigsqcap R).(\bigsqcap N) \in P'$ iff $(R,N) \in succ_\mathcal{T}((\pi_1(t'),\emptyset,\emptyset)) \setminus succ_\mathcal{T}(t')$. Furthermore, by construction of elements in $N_{(\mathcal{T},\mathcal{A})}$, we find $t' \in child_\mathcal{T}(tail(c))$. This means that to perform Rule 6, we only need a second quadruple $(t,P,Q,H)$ for which we have to ensure that $\{s'' \mid \exists r''.s'' \in P'\} \in H$.

So, suppose $\exists r''.s'' \in P'$. Since, by induction hypothesis, $s_{\rho'}(d) \in T_{\mathcal{I}_{|d|}, C_\mathcal{T} \cup s_{\rho'}}(S_n)$, we know $s''(c) \in S_n$. Thus, we can also apply the induction hypothesis to $s''(c)$, to find that $\rho_{s''} = (inv(tail(c)), P_{s''}, Q_{s''}, H_{s''}) \in psat_{C,\mathcal{T}}$ such that $P_{s''}$ and $Q_{s''}$ satisfy the constraints in 1b and 1c of the claim we are currently proving, and $s'' \in H_{s''}$. Furthermore, we find that $s_{\rho_{s''}}(c) \in T_{\mathcal{I}_{|c|}, C_\mathcal{T}}(S_n)$.

For each $\exists r''.s'' \in P'$, we can construct such a quadruple $s_{\rho''}$ and combine them with Rule 2 to conclude

$$\rho'' = (inv(tail(c)), \bigcup_{\exists r''.s'' \in P'} P_{s''}, \bigcup_{\exists r''.s'' \in P'} Q_{s''}, \bigcup_{\exists r''.s'' \in P'} H_{s''}) \in psat_{C,\mathcal{T}},$$

which can serve exactly as the quadruple we were looking for to apply Rule 6, which we also do. Thus, we find that

$$\rho = (inv(tail(c)), \bigcup_{\exists r''.s'' \in P'} P_{s''}, \bigcup_{\exists r''.s'' \in P'} Q_{s''}, \bigcup_{\exists r''.s'' \in P'} H_{s''} \cup \{s\}) \in psat_{C,\mathcal{T}},$$

too. Note that this quadruple satisfies all properties requested in the claim, which naturally follows from the point that all $\rho_{s''}$ satisfy these properties. Furthermore, note that $s_{\rho''}(c) \in T_{\mathcal{I}_{|c|}, C_\mathcal{T} \cup s_{\rho''}}(S_n)$ holds, because for each $\exists r''.s'' \in P'$, we have $s_{\rho_{s''}}(c) \in T_{\mathcal{I}_{|c|}, C_\mathcal{T} \cup s_{\rho_{s''}}}(S_n)$. Thus, we can conclude that $s_\rho(c) \in T_{\mathcal{I}_{|c|}, C_\mathcal{T} \cup s_\rho}(S_n)$, as required.

In case $\{s'' \mid \exists r''.s'' \in P'\} = \emptyset$, the above does not work. Instead, we can simply apply Rule 1 to construct the required $\rho = (t,P,Q,H)$, such that $P = \{\exists(\bigsqcap R).(\bigsqcap N) \mid (R,N) \in succ_\mathcal{T}((\pi_1(t),\emptyset,\emptyset)) \setminus succ_\mathcal{T}(t)\}$ and $Q = \{\exists(\bigsqcap R).(\bigsqcap N) \mid (R,N) \in succ_\mathcal{T}((\pi_1(t),\emptyset,\emptyset))\}$. Then a simple argument suffices to show that $s_\rho(c) \in T_{\mathcal{I}_{|c|}, C_\mathcal{T} \cup s_\rho}(S_n)$, which concludes the induction step. ♣

**Proposition 6.8.** *Given a normalised Horn-$\mathcal{ALCHIQ}$ TBox $\mathcal{T}$ and $C$ a set of positive, normalised constraints, for every target $\mathcal{G}$ and every ABox $\mathcal{A}$ that is consistent with $\mathcal{T}$, we have that if $can_0(\mathcal{T},\mathcal{A})$ validates $(C_\mathcal{T},\mathcal{G})$, then $(\mathcal{T},\mathcal{A})$ validates $(C,\mathcal{G})$.*

**Proof.** We consider the following claim. Taking $i = 0$ suffices to conclude the above proposition.

**Claim 6.9.** *Let $\mathcal{I}_i = can_i(\mathcal{T},\mathcal{A})$. If $\rho = (t,P,Q,H) \in psat_{C,\mathcal{T}}$ with $s \in H$ and $s_\rho(c) \in PA(C_\mathcal{T} \cup \{s_\rho\}, \mathcal{I}_i)$ for some $c \in N_{(\mathcal{T},\mathcal{A})}$ such that $|c| = i$, then $s(c) \in PA(C, can(\mathcal{T},\mathcal{A}))$.*

We show this by letting $i$ decrease step-by-step until reaching 0. Note that because we are considering a least fixed point computation and $C \subseteq C_\mathcal{T}$, there must exist an $i = n$ such that the above naturally holds.

Assume $i = n-1$. Suppose that $s_\rho(c) \in PA(C_\mathcal{T} \cup \{s_\rho\}, \mathcal{I}_{n-1})$, such that $|c| = n-1$. The proof is based on unwinding how $\rho$ got constructed, by an induction on the construction. Without loss of generality, we can assume $s$ got added to $H$ in the last rule applied. Now suppose the last rule applied was Rule 6 (the other cases are rather straightforward). Thus, there exists some $s \leftarrow \exists r.s' \in C$ and moreover we find that $\{(t,P,Q,H \setminus \{s\}), (t',P',Q',H')\} \in psat_{C,\mathcal{T}}$, such that $s' \in H'$. Let $\rho' = (t',P',Q',H')$ and note that because we were able to perform Rule 6, we can conclude that $s'_{\rho'}(ct') \in PA(C_\mathcal{T} \cup \{s'_{\rho'}\}, \mathcal{I}_n)$. By construction of the austere core model, it is clear that in $\mathcal{I}_n$ the node $ct'$ satisfies exactly all concept names mentioned in $\pi_1(t')$; if $\exists(\bigsqcap R).\bigsqcap N \in P'$, it follows that $R \subseteq \pi_2(t')$ and $N \subseteq \pi_3(t')$; and if $\exists r'.s'' \in P$, then $s''$ must be contained in $H$, which means we can first perform this same proof on $s''$ and the quadruple $(t,P,Q,H \setminus \{s\})$ to conclude that $s''(c) \in PA(C, can(\mathcal{T},\mathcal{A}))$, and thus $s''(c) \in PA(C_\mathcal{T}, \mathcal{I}_n)$ by Proposition 6.6. This means that indeed $ct' \in (\exists r'.s'')^{\mathcal{I}_n,\emptyset}$, as required. As $ct'$ will not have any successor nodes in $\mathcal{I}_n$, the conditions posed by $Q'$ are enforced too.

Combining $s'_{\rho'}(ct') \in PA(C_\mathcal{T} \cup \{s'_{\rho'}\}, \mathcal{I}_n)$ with assuming the claim holds for $i = n$, we derive that $s'(ct') \in PA(C, can(\mathcal{T},\mathcal{A}))$. Since $s \leftarrow \exists r.s' \in C$, and $r^- \in \pi_2(t')$ we find $s(c) \in PA(C, can(\mathcal{T},\mathcal{A}))$ too, as required. ♣

These results can be summarised in the following way.

**Theorem 6.10.** *Given a normalised Horn-$\mathcal{ALCHIQ}$ TBox $\mathcal{T}$ and $C$ a set of positive, normalised constraints, for every target $\mathcal{G}$ and every ABox $\mathcal{A}$ that is consistent with $\mathcal{T}$, we have that $(\mathcal{T},\mathcal{A})$ validates $(C,\mathcal{G})$ iff $\mathcal{A}_\mathcal{T}$ validates $(C_\mathcal{T},\mathcal{G})$.*

## 7. Rewriting for stratified SHACL

We now extend the rewriting to constraint sets $C$ with stratified negation. Intuitively, this is done by running a saturation procedure in the rewriting for each stratum of $C$, starting with the lowermost. For the transition to the next stratum in the rewriting procedure, we need to ensure that the outcome of the saturation at a non-topmost stratum is completed with negative information. To this end, we w.l.o.g. assume that all constraints from $C$ with the same shape name on the left-hand-side occur together in the same stratum.

We will operate again on quadruples $(t,P,Q,H)$, which are similar to the ones in the previous section, except that $H$ might additionally contain expressions of the form $\neg s$ for a shape name $s$. For a set $K$ of such quadruples, we say $(t,P,Q,H)$ is *maximal* in $K$ if $(t,P,Q,H) \in K$ and there is no $H' \supset H$ with $(t,P,Q,H') \in K$. Then the notion of *completion* is defined as follows:

**Definition 7.1.** The *completion* $comp_{C,\mathcal{T}}(K)$ of a set $K$ of quadruples w.r.t.a normalised Horn-$\mathcal{ALCHIQ}$ TBox $\mathcal{T}$ and a set of constraints $C$ is a set, defined as follows:

$$comp_{C,\mathcal{T}}(K) = \{(t, P, Q \cup Q', H \cup \bar{H}) \mid (t, P, Q, H) \text{ is maximal in } K\},$$

where

$$\bar{H} := \{\neg s \mid s \text{ occurs in } C, s \notin H\}$$
$$Q' := \{\exists r.s' \mid s \leftarrow \exists r.s' \in C, s \notin H\} \cup \{c \mid s \leftarrow c \in C, s \notin H\}.$$

We need to augment the inference rules of the rewriting procedure for positive constraints with an additional rule to handle constraints of the form $s \leftarrow \neg s'$. Furthermore, Rule 6, is updated to also consider the newly added information to $Q$. Note that using the following updated definition of $sat_{C,\mathcal{T}}(K)$ will give the same results as the previously defined $psat_{C,\mathcal{T}}(K)$ in Definition 6.3 on the first stratum, as the proposed changes do not have any effect if no $\neg\exists r.s$'s or $\neg s$'s are present in the quadruples.

**Definition 7.2.** Given a normalised Horn-$\mathcal{ALCHIQ}$ TBox $\mathcal{T}$ and $C$ a set of normalised constraints, we let $sat_{C,\mathcal{T}}(K)$ be the smallest set of pairs containing $K$ closed under Rules 2–5 defined of Definition 6.3, where '$psat_{C,\mathcal{T}}$' is replaced by '$sat_{C,\mathcal{T}}(K)$', and additionally under the following Rule 6' and 7:

6'. If $s \leftarrow \exists r.s' \in C$, $\{(t, P, Q, H), (t', P', Q', H')\} \subseteq sat_{C,\mathcal{T}}(K)$, $r^- \in \pi_2(t')$, $P' \cap N_I = \emptyset$, $s' \in H'$ and $inv(t') \in child_{\mathcal{T}}(inv(t))$ such that
   - $\exists(\bigsqcap R).\bigsqcap N \in P'$ iff $(R, N) \in succ_{\mathcal{T}}(\pi_1(t'), \emptyset, \emptyset) \setminus succ_{\mathcal{T}}(t')$;
   - $\{s'' \mid \exists r''.s'' \in P'\} \cup \{\neg s'' \mid \exists r''.s'' \in Q', r'' \in \pi_2(t')\} \subseteq H$,

   then $(t, P, Q, H \cup \{s\})$ belongs to $sat_{C,\mathcal{T}}(K)$.
7. If $s \leftarrow \neg s' \in C$ and $(t, P, Q, H) \in sat_{C,\mathcal{T}}(K)$ such that $\neg s' \in H$, then $(t, P, Q, H \cup \{s\})$ belongs to $sat_{C,\mathcal{T}}(K)$.

Now can we define the inference procedure that processes strata from the lowest to the topmost, performing saturation using the updated set of rules at every stratum, interleaved with a computation of the completion in between.

**Definition 7.3.** For a TBox $\mathcal{T}$ and a constraint set $C$ with stratification $C_0, \ldots, C_n$, we let $K_0 = psat_{C_0,\mathcal{T}}$ and for each $0 < i \leq n$

$$K_i = sat_{C_i,\mathcal{T}}(comp_{C_{i-1},\mathcal{T}}(K_{i-1})).$$

We let $C_{\mathcal{T}} = C_{\mathcal{T},K_n}$, where $C_{\mathcal{T},K_n}$ is defined as in Definition 6.4.

**Example 7.4.** Suppose $\mathcal{T} = \{A \sqsubseteq \exists p.B\}$ and consider the following set of constraints $C = C_0 \cup C_1$:

$$C_0 = \{s_C \leftarrow C, \qquad s' \leftarrow \exists p.s_C\}$$
$$C_1 = \{s'' \leftarrow \exists p.\bar{s_C}, \qquad \bar{s_C} \leftarrow \neg s_C, \qquad s \leftarrow s' \wedge s''\}.$$

We consider the following two 2-types in the rest of this example: $t_A := (\{A\}, \emptyset, \emptyset)$ and $t := (\{A\}, \{p\}, \{B\})$. Because of Rule 1, we find that the following two quadruples are contained in $K_0$:

$$(t_A, \{\top\}, \{\exists p.B\}, \emptyset), \qquad (inv(t), \{\top\}, \emptyset, \emptyset).$$

Here, the first one can be updated to $(t_A, \{\top, \exists p.s_C\}, \{\exists p.B\}, \{s'\})$. Note that all three quadruples are maximal in $K_0$, which means we find the following quadruples in $comp_{C_0,\mathcal{T}}$:

$$(t_A, \{\top\}, \{\exists p.B, \exists p.s_C\}, \{\neg s' \neg s_C\}), \qquad (t_A, \{\top, \exists p.s_C\}, \{\exists p.B\}, \{s', \neg s_C\}), \qquad (inv(t), \{\top\}, \{\exists p.s_C\}, \{\neg s', \neg s_C\}).$$

By definition, these quadruples are also contained in $K_1$, and thus, because of Rule 7, also the versions in which $\bar{s_C}$ is added to the fourth set in each quadruple. Then, we can apply Rule 6' to $(t_A, \{\top, \exists p.s_C\}, \{\exists p.B\}, \{s', \neg s_C, \bar{s_C}\})$, $(inv(t), \{\top\}, \{\exists p.s_C\}, \{\neg s', \neg s_C, \bar{s_C}\})$ and the constraint $s'' \leftarrow \exists p.\neg s_C$, to find

$$(t_A, \{\top, \exists p.s_C\}, \{\exists p.B\}, \{s', \neg s_C, \bar{s_C}, s''\}) \in K_1$$

too. After applying Rule 5 based on the constraint $s \leftarrow s' \wedge s'' \in C_1$, we can extract the following rewritten constraint:

$$s \leftarrow A \wedge \neg B \wedge \neg C \wedge \exists p.s_C \wedge \neg\exists p.B \in C_{\mathcal{T}},$$

Now let $\mathcal{A} = \{A(a), p(a, b), C(b)\}$ and set the target to $\mathcal{G} = \{s(a)\}$. As $\mathcal{A} = \mathcal{A}_{\mathcal{T}}$, we indeed find a positive answer when validating $(C_{\mathcal{T}}, \mathcal{G})$ over $\mathcal{A}_{\mathcal{T}}$. Notice $can(\mathcal{T}, \mathcal{A})$ corresponds to the canonical interpretation of $\{A(a), p(a, b), C(b), p(a, at), B(at)\}$. Thus, we can conclude that $(\mathcal{T}, \mathcal{A})$ validates $(C, \mathcal{G})$ too. ♡

Given a stratified set of constraints $C$ with stratification $C_0, \ldots, C_n$, let $C_{\leq i} := C_0 \cup \ldots \cup C_i$. Similarly, we will use $C_{\mathcal{T},\leq i}$ to denote $C_{\mathcal{T},K_0} \cup \ldots \cup C_{\mathcal{T},K_i}$.

**Theorem 7.5.** *Given a normalised Horn-$\mathcal{ALCHIQ}$ TBox $\mathcal{T}$ and $C$ a set of stratified constraints with stratification $C_0, \ldots, C_n$, for every target $\mathcal{G}$ and every ABox $\mathcal{A}$ that is consistent with $\mathcal{T}$, we have that $(\mathcal{T}, \mathcal{A})$ validates $(C, \mathcal{G})$ iff $\mathcal{A}_{\mathcal{T}}$ validates $(C_{\mathcal{T}}, \mathcal{G})$.*

**Proof.** (⇒). We will prove completeness by proving the following claim by induction on $i$

**Claim 7.6.** *Let $\mathcal{I}_i = can_i(\mathcal{T}, \mathcal{A})$. For each $c \in N_{(\mathcal{T},\mathcal{A})}$ and for each $0 \leq i \leq n$, there exists $\rho = (t, P, Q, H) \in K_i$ that is maximal in $K_i$, and moreover:*

1. *if* $c \in N_{(\mathcal{T},\mathcal{A})} \setminus N_I$, *then*
   (a) $t = inv(tail(c))$;
   (b) $P \supseteq \{\exists(\bigsqcap R).\bigsqcap N \mid (R,N) \in succ_{\mathcal{T}}((\pi_1(t),\emptyset,\emptyset)) \setminus succ_{\mathcal{T}}(t)\}$ *and* $P \cap N_I = \emptyset$;
   (c) $Q \supseteq \{\exists(\bigsqcap R).\bigsqcap N \mid (R,N) \in succ_{\mathcal{T}}((\pi_1(t),\emptyset,\emptyset))\}$;
2. *for all other* $c \in N_I$, *we have*
   (a) $t = (\{A \in N_C \mid A(c) \in \mathcal{A}_{\mathcal{T}}\},\emptyset,\emptyset)$;
   (b) $P \supseteq \{\exists(\bigsqcap R).\bigsqcap N \mid (R,N) \in succ_{\mathcal{T}}(t), c \in (\exists(\bigsqcap R).\bigsqcap N)^{\mathcal{A}_{\mathcal{T}}}\}$ *and* $(P \cap N_I) \setminus \{c\} = \emptyset$;
   (c) $Q \supseteq \{\exists(\bigsqcap R).\bigsqcap N \mid (R,N) \in succ_{\mathcal{T}}(t), c \notin (\exists(\bigsqcap R).\bigsqcap N)^{\mathcal{A}_{\mathcal{T}}}\}$;
3. $H = \{s \mid s(c) \in PA(can(\mathcal{T},\mathcal{A}),\mathcal{C}_{\leq i})\} \cup \{\neg s \mid s(c) \notin PA(can(\mathcal{T},\mathcal{A}),\mathcal{C}_{\leq i}), s$ *defined in* $\mathcal{C}_{i-1}\}$;
4. *and* $s_\rho(c) \in PA(\mathcal{I}_{|c|},\mathcal{C}_{\mathcal{T},\leq i} \cup s_\rho)$.

Analogous to Proposition 6.6, it follows that for each $s$ such that $s(c) \in PA(can(\mathcal{T},\mathcal{A}),\mathcal{C}_{\leq 0})$, there exists a quadruple $(t,P_s,Q_s,H_s) \in sat_{\mathcal{C}_0,\mathcal{T}}(\emptyset) = K_0$ such that $s \in H_s$ and such that requirements 1, 2 and 4 of the claim hold. Now let $X = \{s \mid s(c) \in PA(can(\mathcal{T},\mathcal{A}),\mathcal{C}_{\leq 0})\}$. Then, following Rule 2 of the rewriting algorithm in Definition 6.3, we must conclude that also $(t,\bigcup_{s\in X} P_s,\bigcup_{s\in X} Q_s,\bigcup_{s\in X} H_s) \in K_0$. This quadruple is exactly the quadruple we were looking for: requirements 1, 2 and 4 have to be satisfied, as they are satisfied for each $(t,P_s,Q_s,H_s)$ separately; requirement 3 follows, as $s \in H_s$ implies $X \subseteq \bigcup_{s\in X} H_s$, and at the same time the correctness of the rewriting forbids that $H_s \setminus X \neq \emptyset$. In the same way, maximality of the quadruple is implied.

For the induction step, $i = j+1$, note that for each $c \in N_I \cup N_{(\mathcal{T},\mathcal{A})}$, we are given some $(t,P,Q,H) \in K_j$ that is maximal and satisfies requirements 1 to 4 for $i = j$. As this quadruple is assumed to be maximal, completion $comp_{\mathcal{C}_j,\mathcal{T}}$ can be applied to $(t,P,Q,H)$. Thus, we find $(t,P,Q\cup Q',H\cup\bar{H}) \in K_{j+1}$. Since $s \in H$ iff $s(c) \in PA(can(\mathcal{T},\mathcal{A}),\mathcal{C}_{\leq j})$, it must follow that $\neg s \in \bar{H}$ iff $s(c) \notin PA(can(\mathcal{T},\mathcal{A}),\mathcal{C}_{\leq j})$ and $s$ occurs in $\mathcal{C}_{\leq j}$. Now suppose $s(c) \in PA(can(\mathcal{T},\mathcal{A}),\mathcal{C}_{\leq j+1}) \setminus PA(can(\mathcal{T},\mathcal{A}),\mathcal{C}_{\leq j})$, then there must exist a constraint $s \leftarrow C \in \mathcal{C}$ that caused this addition.

If $C = \neg s'$, then $s'$ is defined in $\mathcal{C}_{\leq j}$ by the stratification rules, and $s'(c) \notin PA(can(\mathcal{T},\mathcal{A}),\mathcal{C}_{\leq j})$, which means $\neg s' \in \bar{H}$, thus Rule 7 can be applied to conclude that $(t,P,Q\cup Q',H\cup\bar{H}\cup\{s\}) \in K_{j+1}$. If $C$ is any of the other options, the same reasoning as in the proof of Proposition 6.6 can be used.

Now all that is left to show, is that each $Y \in Q'$ is not blocking the satisfaction of requirement 4. By taking a closer look a the completion procedure, we see that $Y = \exists r.s'$ for some $s \leftarrow \exists r.s' \in \mathcal{C}_{\leq j}$ such that $s \notin H$, or $Y = c$, for some $s \leftarrow c \in \mathcal{C}_{\leq j}$ and $s \notin H$. For both cases, note that $s \notin H$, which means $s(c) \notin PA(can(\mathcal{T},\mathcal{A}),\mathcal{C}_{\leq j})$. Assuming that any of these $Y$ is indeed blocking restriction 4, directly leads to a contradiction with the previous observation. Thus, we have indeed found the required quadruple.

($\Leftarrow$). Again, let $\mathcal{I}_i = can_i(\mathcal{T},\mathcal{A})$. To show correctness, we will show by induction on $i$ that for each $c \in N_{(\mathcal{T},\mathcal{A})}$ and for each $\rho = (t,P,Q,H) \in K_i$ such that (a) $s_\rho(c) \in PA(\mathcal{I}_{|c|},\mathcal{C}_{\mathcal{T},\leq i} \cup s_\rho)$, we find for each $s \in H$ that $s(c) \in PA(can(\mathcal{T},\mathcal{A}),\mathcal{C}_{\leq i})$.

First, note that Proposition 6.8 provides the induction basis. So, for the induction step, suppose we are given some $c \in N_{(\mathcal{T},\mathcal{A})}$ and $(t,P,Q,H) \in K_{i+1}$ such that (a) holds. We will focus on the case that there exists a $(t,P,Q',H') \in K_i$ that is maximal in $K_i$, such that $(t,P,Q,H)$ is the result of first applying completion to $(t,P,Q',H')$ to find $(t,P,Q'\cup Q'',H'\cup\bar{H}') \in comp_{\mathcal{C}_i,\mathcal{T}}$, followed by application(s) of rules 3–5, 6' or 7. Here, we will only consider the result of applying Rule 7, as the treatment of the application of the other rules is very similar to what is discussed in Proposition 6.8. Note that Rules 1 and 2 are not really having any relevant effects after finishing the first stratum.

So, assume $s \leftarrow \neg s' \in \mathcal{C}_{i+1}$ and $\neg s' \in \bar{H}'$. We argue that each $s' \leftarrow C \in \mathcal{C}$ cannot be used to deduce $s'(c) \in PA(can(\mathcal{T},\mathcal{A}),\mathcal{C}_{\leq i})$.

- $C \in N_C$. If $C \in \pi_1(t)$, we would have found $s' \in H'$ by maximality of $(t,P,Q',H')$. Thus, $C \notin \pi_1(t)$, which means $\neg C$ appears in $s_{(t,P,Q,H)}$. Since (a) holds, this means that $c \notin C^{can(\mathcal{T},\mathcal{A})}$, which suffices.
- $C \in N_I$. As $\neg s' \in \bar{H}'$, we find $C \in Q$. Since (a) holds, this means $C^{can(\mathcal{T},\mathcal{A})} \neq c^{can(\mathcal{T},\mathcal{A})}$, which suffices.
- $C \in \{s'', s''\wedge s''' \mid \{s'',s'''\} \subseteq N_S\}$. Perform this same argument for $s''$ or $s'''$. As we are considering a least fixed point semantics, this regression must terminate.
- $C = \exists r.s''$, for some $r \in N_R$, $s \in N_S$. As $\neg s' \in \bar{H}'$, we find $C \in Q$. Note we derived completeness and correctness of all strata up till $i$, that is for all $s$, for all natural numbers $n$, and all $c$ such that $|c| \leq n$, we have $s(c) \in PA(can(\mathcal{T},\mathcal{A}),\mathcal{C}_{\leq i})$ iff $s(c) \in PA(\mathcal{I}_n,\mathcal{C}_{\mathcal{T},\leq i})$. Thus, combining this with $(a)$, we find that in any case $s''(d) \notin PA(can(\mathcal{T},\mathcal{A}),\mathcal{C}_{\leq i})$, for $dt = c$, such that $r^- \in \pi_2(t)$.

  The other option we have to exclude is $d = ct$, for some $r \in \pi_2(t)$. For contradiction, assume $s''(d) \in PA(can(\mathcal{T},\mathcal{A}),\mathcal{C}_{\leq i})$. So we know there must exist a quadruple $\rho_d = (inv(t),P_d,Q_d,H_d) \in K_i$ such that $s'' \in H_d$ and $s_{\rho_d}(d) \in PA(\mathcal{I}_{|d|},\mathcal{C}_{\leq i,\mathcal{T}} \cup s_{\rho_d})$, and such that these there does not exist $(inv(t),P'_d,Q'_d,H'_d) \in K_i$ with the same properties, but $|\{\exists r.s \in P'_d\}| < |\{\exists r.s \in P_d\}|$. First of all, note that $\{\neg s \mid \exists r.s \in Q_d\} \subseteq H$; for all such $s$, we have $\neg s \notin H$, implies $s \in H$. Thus, $s(c) \in PA(can(\mathcal{T},\mathcal{A}),\mathcal{C}_{\leq i})$ by induction hypothesis. Using completeness and correctness, this implies $s_{\rho_d}(d) \notin PA(can_{|d|}(\mathcal{T},\mathcal{A}),\mathcal{C}_{\leq i,\mathcal{T}} \cup s_{\rho_d})$, which is a contradiction.

  Now there are two options: if $\{s \mid \exists r.s \in P_d\} \subseteq H$, then Rule 6' could have been applied on $s' \leftarrow \exists r.s''$ at a certain point, which is in contradiction with the maximality of $(t,P,Q',H')$. So, we are left with the second option: $\{s \mid \exists r.s \in P_d\} \nsubseteq H$. Now we can repeat this same argument for each member of $\{s \mid \exists r.s \in P_d\} \setminus H$. Because we are solely considering the least amount of $\exists r.s$'s in $P_d$, we know they all are essential in deriving $s''(d)$. Thus, $s''(d) \notin PA(can(\mathcal{T},\mathcal{A}),\mathcal{C}_{\leq i})$ for all $d$ such that $(c,d) \in (r)^{can(\mathcal{T},\mathcal{A})}$, which automatically implies the required result.
- $C = \neg s''$. If $s \leftarrow \neg s' \in \mathcal{C}_{i+1}$, then $s' \leftarrow \neg s'' \in \mathcal{C}_{\leq i}$, by definition of the strata. This means that $\neg s''$ cannot be part of $\bar{H}'$. So, to apply Rule 7, it must be that $\neg s'' \in H'$. However, this is in contradiction with the maximality of $(t,P,Q',H')$ that would imply $s' \in H'$, which concludes this case.

Thus, we conclude that $s'(c) \notin PA(can(\mathcal{T},\mathcal{A}),\mathcal{C}_{\leq i})$, and thus that indeed $s(c) \in PA(can(\mathcal{T},\mathcal{A}),\mathcal{C}_{\leq i+1})$, as required. ♣

## 8. Pure rewritings

Theorem 7.5 lets us reduce SHACL validation in the presence of Horn-$\mathcal{ALCHI}$ TBoxes to plain SHACL validation over the completed ABox $\mathcal{A}_\mathcal{T}$. In this section we discuss how to update this rewriting to a pure rewriting. That is, how to produce a set of constraints $C'$, in a data-independent way, such that $(\mathcal{T}, \mathcal{A})$ validates $(C, \mathcal{G})$ iff $\mathcal{A}$ validates $(C', \mathcal{G})$.

To make the rewriting pure, we have to encode the information added to $\mathcal{A}$ by the immediate consequence operator $T_\mathcal{T}$ defined in Definition 3.7 in new constraints. However, SHACL as introduced in [1] can only propagate information on nodes, and information about the properties/roles connecting the nodes cannot be directly represented.

We discuss two ways around this. First we present a solution for TBoxes that do not contain counting axioms ($A \sqsubseteq \leq_1 r.B$). The role information added to $\mathcal{A}_\mathcal{T}$ in this case is simple, and it can be directly derived from locally satisfying axioms of the form $r_0 \sqcap \ldots \sqcap r_n \sqsubseteq r$. For the general case, we extend SHACL and enable it to describe edges, rather than just nodes.

Without loss of generality, we assume in this section that there are no cycles over role inclusions. That is, no TBox contains a set of role inclusions of the form $\{r \sqsubseteq r_1, r_1 \sqsubseteq r_2, \ldots, r_n \sqsubseteq r\}$. If such a cycle exists, then all roles names appearing in the cycle are replaced by one unique role name.

### 8.1. Rewriting for normalised Horn-$\mathcal{ALCHI}$

Let us first focus on how to define a pure rewriting for normalised Horn-$\mathcal{ALCHI}$.

**Definition 8.1.** A Horn-$\mathcal{ALCHI}$ TBox $\mathcal{T}$ is in normal form if each of the concept inclusions in $\mathcal{T}$ are of one of the following forms:

(F1) $A_0 \sqcap \ldots \sqcap A_n \sqsubseteq B$ (F3) $A \sqsubseteq \forall r.B$

(F4) $A \sqsubseteq \exists r.B$,

for $\{A, A_0, \ldots, A_n, B\} \subseteq \overline{N_C}$ and $r \in \overline{N_R}$. Furthermore, $\mathcal{T}$ may contain role inclusions of the form $r \sqsubseteq r'$, for $\{r, r'\} \subseteq \overline{N_R}$.

For each Horn-$\mathcal{ALCHI}$ TBox $\mathcal{T}$, we define the following set of constraints, capturing most of the information derived by the immediate consequence operator $T_\mathcal{T}$, defined in Definition 3.7.

**Definition 8.2.** For each $A \in N_C$, we let $s_A \in N_S$ be a fresh shape name. Given a normalised Horn-$\mathcal{ALCHI}$ TBox $\mathcal{T}$, let $\mathcal{T}_s$ be the smallest set of constraints containing for each $\mathcal{T} \models A_0 \sqcap \ldots \sqcap A_n \sqsubseteq B$, the constraint

$$s_B \leftarrow \bigwedge_{0 \leq i \leq n} s_{A_i} \in \mathcal{T}_s, \tag{2}$$

furthermore, for each $A \sqsubseteq \forall r.B \in \mathcal{T}$ and $\mathcal{T} \models S \sqsubseteq r$ the constraint

$$s_B \leftarrow \exists S^-.s_A \in \mathcal{T}_s, \tag{3}$$

and for each $A \in \overline{N_C}$, the constraint

$$s_A \leftarrow A \in \mathcal{T}_s. \tag{4}$$

**Definition 8.3.** Given a normalised Horn-$\mathcal{ALCHI}$ TBox $\mathcal{T}$, let $C^+_\mathcal{T}$ be the set of constraints consisting of the constraints in $C_\mathcal{T}$, taking into account the following alterations:

- each concept name $A \in N_C$ appearing in an axiom in $C_\mathcal{T}$ is replaced by $s_A$; and
- for each $\mathcal{T} \models \bigsqcap R \sqsubseteq r$, such that $R \cup \{r\} \subseteq R'$, and $\bigsqcap R'$ appears in some constraint, $\bigsqcap R'$ is replaced by $\bigsqcap(R' \setminus \{r\})$.

Note that when considering $C^+_\mathcal{T} \cup \mathcal{T}_s$ instead of $C_\mathcal{T}$, we are introducing a new lowest stratum: the set of freshly introduced constraints, based on existing concept names. However, the only point of this stratum is to mimic exactly which concept names hold at which points in the enriched ABox $\mathcal{A}_\mathcal{T}$, defined in Definition 3.7.

**Proposition 8.4.** *Given a normalised Horn-$\mathcal{ALCHI}$ TBox $\mathcal{T}$ and $C$ a set of positive constraints, for every target $\mathcal{G}$ and every ABox $\mathcal{A}$ that is consistent with $\mathcal{T}$, we have that $can_0(\mathcal{T}, \mathcal{A})$ validates $(C_\mathcal{T}, \mathcal{G})$ iff $\mathcal{A}$ validates $(C^+_\mathcal{T} \cup \mathcal{T}_s, \mathcal{G})$.*

**Proof.** Note that $can_0(\mathcal{T}, \mathcal{A})$ corresponds to $\mathcal{A}_\mathcal{T}$, the first layer in building the austere canonical model, defined in Definition 3.7. In this definition, the ABox is completed in a least fixed point manner under certain axioms ($A_0 \sqcap \ldots \sqcap A_n \sqsubseteq B$ and $A \sqsubseteq \forall r.B$) that are precisely encoded in constraints of the form (2) and (3). The only difference is that we cannot translate axioms of the form $\mathcal{T} \models S \sqsubseteq r'$ directly. A simple fix suffices: as role atoms in $\mathcal{A}_\mathcal{T} \setminus \mathcal{A}$ can only be the result of an axiom of the form $r \sqsubseteq r'$, the replacement in Definition 8.3 in combination with the implied roles in constraints of the form (3) is sufficient. As we are also considering a least fixed point semantics for SHACL validation, the above result directly follows. ♣

Thus, considering $C^+_\mathcal{T} \cup \mathcal{T}_s$ as rewritten set of constraints indeed provides us with a pure rewriting. The following theorem follows from combining the results of Theorem 7.5 and Proposition 8.4.

**Theorem 8.5.** *Given a normalised Horn-$\mathcal{ALCHI}$ TBox $\mathcal{T}$ and $C$ a set of stratified constraints, for every target $\mathcal{G}$ and every ABox $\mathcal{A}$ that is consistent with $\mathcal{T}$, we have that $(\mathcal{T}, \mathcal{A})$ validates $(C, \mathcal{G})$ iff $\mathcal{A}$ validates $(C^+_\mathcal{T}, \mathcal{G})$.*

### 8.2. $SHACL^b$

For a pure rewriting of normalised Horn-$\mathcal{ALCHIQ}$, we propose a different solution: we define an extension of SHACL that we call SHACL$^b$. Here we also allow 'binary' shape constraints $b$: we can also express constraints on pairs of nodes. We note that not all operators defined in SHACL$^b$ are necessary to define a pure rewriting for normalised Horn-$\mathcal{ALCHIQ}$. However, we do resort to the following version to demonstrate how the imbalance between unary and binary (shape) atoms in SHACL can be fully reduced.

**Definition 8.6.** Let *SHACL*$^b$ consist of shape constraints of the form $s \leftarrow \varphi$, or $b \leftarrow \psi$, for $\{s, b\} \subseteq N_S$, such that

$$\begin{aligned}\varphi &::= c \mid s \mid A \mid \varphi \wedge \varphi \mid \neg\varphi \mid \exists\psi.\varphi\\ \psi &::= s? \mid b \mid r \mid \psi \cup \psi \mid \psi \cap \psi \mid \psi \cdot \psi \mid \psi^* \mid \psi^- \mid \psi \setminus \psi,\end{aligned}$$

where $c \in N_I$, $\{s, b\} \subseteq N_S$, $A \in N_C$ and $r \in \overline{N_R}$. We evaluate SHACL$^b$ as defined in Fig. 4, extended with

$$\begin{aligned}(\exists\psi.\varphi)^{\mathcal{I},S} &:= \{e \in \Delta^{\mathcal{I}} \mid \exists e' \in \Delta^{\mathcal{I}} : (e, e') \in (\psi)^{\mathcal{I},S} \wedge e' \in (\varphi)^{\mathcal{I},S}\}\\ (s?)^{\mathcal{I},S} &:= \{(e, e) \in \Delta^{\mathcal{I}} \times \Delta^{\mathcal{I}} \mid e \in s^{\mathcal{I},S}\}\\ (b)^{\mathcal{I},S} &:= \{(e, e') \in \Delta^{\mathcal{I}} \times \Delta^{\mathcal{I}} \mid b(e, e') \in S\}\\ (r)^{\mathcal{I},S} &:= r^{\mathcal{I}}\\ (\psi \cup \psi')^{\mathcal{I},S} &:= (\psi)^{\mathcal{I},S} \cup (\psi')^{\mathcal{I},S}\\ (\psi \cap \psi')^{\mathcal{I},S} &:= (\psi)^{\mathcal{I},S} \cap (\psi')^{\mathcal{I},S}\\ (\psi \cdot \psi')^{\mathcal{I},S} &:= \{(e, e') \in \Delta^{\mathcal{I}} \times \Delta^{\mathcal{I}} \mid \exists e'' \in \Delta^{\mathcal{I}} : (e, e'') \in (\psi)^{\mathcal{I},S} \wedge (e'', e') \in (\psi')^{\mathcal{I},S}\}\\ (\psi^*)^{\mathcal{I},S} &:= \{(e, e') \in \Delta^{\mathcal{I}} \times \Delta^{\mathcal{I}} \mid \exists\{e_0, \dots, e_n\} \subseteq \Delta^{\mathcal{I}} :\\ &\qquad \forall i \in \{0, \dots, n_1\} : (e_i, e_{i+1}) \in (\psi)^{\mathcal{I},S} \wedge e_0 = e \wedge e_n = e'\}\\ (\psi^-)^{\mathcal{I},S} &:= \{(e, e') \in \Delta^{\mathcal{I}} \times \Delta^{\mathcal{I}} \mid (e', e) \in (\psi)^{\mathcal{I},S}\}\\ (\psi \setminus \psi')^{\mathcal{I},S} &:= (\psi)^{\mathcal{I},S} \setminus (\psi')^{\mathcal{I},S}.\end{aligned}$$

The following shape expressions, although not explicitly present, are also expressible in SHACL$^b$.

$$\begin{aligned}\exists R.\varphi &= \exists(r_0 \cap \dots \cap r_n).\varphi\\ \mathsf{eq}(E, E') &= \neg\exists.(E \setminus E').\top \wedge \neg\exists.(E' \setminus E).\top\\ \mathsf{disj}(E, E') &= \neg\exists(E \cap E').\top\end{aligned}$$

where $R = r_0 \sqcap \dots \sqcap r_n$.

**Definition 8.7.** Define $s_A \in N_S$ and $b_r \in N_S$ to be a fresh shape names for each $A \in N_C$, respectively $r \in \overline{N_R}$. Given a normalised Horn-$\mathcal{ALCHIQ}$ TBox $\mathcal{T}$, let $\mathcal{T}_s$ be the smallest set of binary SHACL constraints containing for each $\mathcal{T} \models A_0 \sqcap \dots \sqcap A_n \sqsubseteq B$, the constraint

$$s_B \leftarrow \bigwedge_{0 \le i \le n} s_{A_i} \in \mathcal{T}_s, \tag{5}$$

for each $A \sqsubseteq \forall r'.B \in \mathcal{T}$ the constraint

$$s_B \leftarrow \exists b_{r^-}.s_A \in \mathcal{T}_s, \tag{6}$$

and, for each $\mathcal{T} \models A_0 \sqcap \dots \sqcap A_n \sqsubseteq \exists(r_0 \sqcap \dots \sqcap r_m).(B_0 \sqcap \dots \sqcap B_k)$ and $A \sqsubseteq \le_1 r.B \in \mathcal{T}$ such that $r \in \{r_0, \dots, r_m\}$ and $B \in \{B_0, \dots, B_k\}$, the constraints

$$\{b_{r_i} \leftarrow \hat{s}? \cdot b_r \cdot s_B?, \qquad \hat{s} \leftarrow s_A \wedge s_{A_0} \wedge \dots \wedge s_{A_n}, \tag{7}$$

$$s_{B_j} \leftarrow s_B \wedge \exists b_{r^-}.\hat{s}\} \subseteq \mathcal{T}_s, \tag{8}$$

for each $i \in \{0, \dots, m\}$ and $j \in \{0, \dots, k\}$, and a fresh shape name $\hat{s}$, furthermore, for each $r \sqsubseteq r' \in \mathcal{T}$, the constraint

$$b_{r'} \leftarrow b_r \in \mathcal{T}_s, \tag{9}$$

and for each $r \in \overline{N_R}$, $p \in N_R$ the constraints

$$\{b_r \leftarrow r, \qquad b_{(p^-)} \leftarrow (b_p)^-, \qquad b_p \leftarrow (b_{p^-})^-\} \subseteq \mathcal{T}_s, \tag{10}$$

and, lastly, for each $A \in \overline{N_C}$, the constraint

$$s_A \leftarrow A \in \mathcal{T}_s. \tag{11}$$

**Proposition 8.8.** *Given a normalised Horn-$\mathcal{ALCHIQ}$ TBox $\mathcal{T}$, for each $\mathcal{A}$, we find that for all $A \in N_C$, all $r \in \overline{N_R}$, and all $\{c, d\} \subseteq \Delta^{\mathcal{A}}$,*

- *$A(c) \in \mathcal{A}_{\mathcal{T}}$ iff $s_A(c) \in PA(\mathcal{T}_s, \mathcal{A})$;*
- *$r(c, d) \in \mathcal{A}_{\mathcal{T}}$ iff $b_r(c, d) \in PA(\mathcal{T}_s, \mathcal{A})$.*

**Proof.** Let $\mathcal{T}_s^*$ consists of the constraints in $\mathcal{T}_s$ defined by (10) and (11). Then, it is clear that $A(c) \in \mathcal{A}$ iff $s_A(c) \in PA(\mathcal{T}_s^*, \mathcal{A})$ and $r(c,d) \in \mathcal{A}$ iff $b_r(c,d) \in PA(\mathcal{T}_s^*, \mathcal{A})$. This means that all information of the ABox can be precisely captured in shape names. For the rest, note how the different lines in Definition 3.7 defining the immediate consequence operator used to build $\mathcal{A}_\mathcal{T}$ from $\mathcal{A}$ correspond one-to-one to the constraints defined in (5) to (9), with the exception of using the shorthand $\hat{s}$. As both the perfect assignment $PA$ as the construction of $\mathcal{A}_\mathcal{T}$ are based on a least fixed point semantics, the result follows. ♣

The completeness and correctness of the full rewriting now directly follow. Clearly, we concluded that all information in $\mathcal{A}_\mathcal{T}$, albeit in the form $s_A$ and $b_r$, can be derived by SHACL$^b$ constraints constructed to this end. Therefore, there is not really a difference between evaluating $C_\mathcal{T}$ over $\mathcal{A}_\mathcal{T}$, or $C_\mathcal{T} \cup \mathcal{T}_s$ over $\mathcal{A}$, except that we should replace concept names and roles by their shape names referring to them.

**Definition 8.9.** Given a normalised Horn-$\mathcal{ALCHIQ}$ TBox $\mathcal{T}$, let $C_\mathcal{T}^+$ be the set of constraints consisting of the axioms in $C_\mathcal{T}$, such that each concept name $A \in N_C$ appearing in an axiom in $C_\mathcal{T}$ is replaced by $s_A$, and similarly, each $r \in \overline{N_R}$ by $b_r$.

Now that the concept and role names are replaced by their shape equivalents, we have indeed produced a pure rewriting.

**Theorem 8.10.** *Given a normalised Horn-$\mathcal{ALCHIQ}$ TBox $\mathcal{T}$ and $C$ a set of stratified constraints, for every target $\mathcal{G}$ and every ABox $\mathcal{A}$ that is consistent with $\mathcal{T}$, we have that $(\mathcal{T}, \mathcal{A})$ validates $(C, \mathcal{G})$ iff $\mathcal{A}$ validates $(C_\mathcal{T}^+ \cup \mathcal{T}_s, \mathcal{G})$.*

## 9. Complexity results

We now discuss the computational complexity of SHACL validation in the presence of Horn-$\mathcal{ALCHIQ}$ TBoxes. Specifically, we discuss the *combined complexity* and the *data complexity* of the problem [34]. The former is measured in terms of the combined size of all input components, while the latter is measured assuming all components except the ABox are of fixed size.

**Theorem 9.1.** *The problem of SHACL validation in the presence of normalised Horn-$\mathcal{ALCHIQ}$ TBoxes is ExpTime-complete in combined complexity and PTime-complete in data complexity.*

**Proof.** For data complexity, the PTime lower bound was shown in [21], which already applies in the absence of ontologies. The matching PTime upper bound follows from Theorem 7.5, combined with the fact that $\mathcal{A}_\mathcal{T}$ has size polynomial in $\mathcal{A}$'s size itself, and that validation under stratified constraints without ontologies in feasible in polynomial time in data complexity [21]. We note that checking whether the input graph is consistent with a TBox can also be done in polynomial time.

For the upper bound in combined complexity, we rely on the rewriting algorithm discussed in Sections 6 and 7. Let $(\mathcal{T}, \mathcal{A})$ be a normalised Horn-$\mathcal{ALCHIQ}$ knowledge base and $C$ any set of (stratified) constraints. Observe that the number of different quadruples $(t, P, Q, H)$ over the signature of $\mathcal{T}$ and $C$ that can be added to any $K_i$ during the rewriting is bounded by an exponential in the size of $\mathcal{T}$ and $C$. An application of any of the rules takes polynomial time in the size of $\mathcal{T}$, $\mathcal{A}$, $C$ and $\bigcup_{i>0} K_i$, which means the application of all rules in the procedure is bounded by an exponential in the size of $\mathcal{T}$, $\mathcal{A}$ and $C$. Computing $C_{\mathcal{T},K}$ is polynomial in the size of $K$, for all possible $K$. Thus, overall we get a procedure that runs in exponential time in the size of $\mathcal{T}$, $\mathcal{A}$ and $C$.

The hardness result in combined complexity follows from the hardness of reasoning in the ontology alone: checking whether an atom of the form $r(a,b)$ or $A(a)$ is implied by a knowledge base is already ExpTime-complete in combined complexity [11]. ♣

The hardness inherited from the ontology alone is not the only source of complexity. ExpTime-hardness holds even for very weak description logics like *DL-Lite*$_\mathcal{R}$ and a very simple fragment of SHACL.

Define a *simple* shape expression $\varphi$ in the following way

$$\varphi ::= s \mid A \mid \varphi \wedge \varphi \mid \exists r.\varphi,$$

for $s \in N_S$, $A \in N_C$ and $r \in \overline{N_R}$. Shape constraints in *simple*-SHACL are then build in the same way as for regular SHACL by using simple shape expressions instead of regular ones. All other definitions extend in a straightforward way to simple-SHACL.

**Theorem 9.2.** *Let $\mathcal{L}$ be any DL that support axioms of the forms $A \sqsubseteq \exists r.\top$ and $\exists r_1.\top \sqsubseteq \exists r_2.\top$ and $r_1 \sqsubseteq r_2$. In the presence of $\mathcal{L}$ TBoxes, simple-SHACL validation is ExpTime-complete in combined complexity.*

**Proof.** Membership follows directly from Theorem 9.1. To prove ExpTime-hardness in combined complexity we reduce the word problem of polynomially space-bounded *Alternating Turing Machines (ATMs)* to validating shapes under an $\mathcal{L}$ ontology.

An *ATM* is defined as a tuple of the form

$$\mathcal{M} = (\Sigma, Q_\exists, Q_\forall, q_0, q_{acc}, q_{rej}, \delta)$$

where $\Sigma$ is an *alphabet*, $Q_\exists$ is a set of *existential states*, $Q_\forall$ is a set of *universal states*, disjoint from $Q_\exists$, $q_0 \in Q_\exists$ is an *initial state*, $q_{acc} \in Q_\exists \cup Q_\forall$ is an *accepting state*, $q_{rej} \in Q_\exists \cup Q_\forall$ is a *rejecting state*, and $\delta$ is a *transition relation* of the form

$$\delta \subseteq Q \times (\Sigma \cup \{B\}) \times Q \times (\Sigma \cup \{B\}) \times \{-1, 0, +1\}$$

with $Q = Q_\exists \cup Q_\forall$. Here $B$ is the *blank symbol*. We let $\delta(q,a) = \{(q', b, D) \mid (q, a, q', b, D) \in \delta\}$. W.l.o.g., we assume that in $\mathcal{M}$ universal and existential states are strictly alternating: if $(q, a, q', b, m) \in \delta$ and $q \in Q_\exists$ (resp., $q \in Q_\forall$), then $q' \in Q_\forall$ (resp., $q' \in Q_\exists$). We further assume that $|\delta(q,a)| = 2$ for all combinations of states $q \in Q$ and symbols $a \in \Sigma \cup \{B\}$. If $\delta(q,a) = \{(q_1, a_1, D_1), (q_2, a_2, D_2)\}$, we let $\delta_\ell(q,a) = (q_1, a_1, D_1)$ and $\delta_r(q,a) = (q_2, a_2, D_2)$.

A run of an ATM $\mathcal{M}$ on an input word $w$ is defined as usual. We assume a word $w = d_1 \cdots d_n \in \Sigma^*$ with $n > 0$ together with an ATM $\mathcal{M}$ that only uses the tape cells where the input word was written, i.e., it only uses the first $n$ cells. Checking if such $\mathcal{M}$ accepts $w$ is an ExpTime-hard problem.

We show how to construct a knowledge base $(\mathcal{T}, \mathcal{A})$ and shapes graph $(C, \{s_{acc}(a)\})$ such that $\mathcal{M}$ accepts $w$ iff $(\mathcal{T}, \mathcal{A})$ validates $(C, \{s_{acc}(a)\})$. The reduction takes polynomial time in the size of $\mathcal{M}$ and $w$. It uses the following symbols:

- a concept name $Init$ and a shape name $s_{acc}$;
- role names $succ, succ_\ell, succ_r$;
- shape names $s_q$ for all states $q \in Q_\exists \cup Q_\forall$;
- shape names $h_i$ for all $1 \le i \le n$;
- shape names $c_b^{(i)}$ for all $b \in \Sigma \cup \{B\}$ and $1 \le i \le n$.

Set $\mathcal{A} = \{Init(a)\}$ and let $\mathcal{T}$ contain the following inclusions:

$$Init \sqsubseteq \exists succ_\ell.\top \qquad succ_\ell \sqsubseteq succ \qquad \exists succ^-.\top \sqsubseteq \exists succ_\ell.\top$$
$$Init \sqsubseteq \exists succ_r.\top \qquad succ_r \sqsubseteq succ \qquad \exists succ^-.\top \sqsubseteq \exists succ_r.\top.$$

The interpretation $can(\mathcal{T}, \mathcal{A})$ provides us with an infinite binary tree. In that tree, the root is representing the starting configuration of $\mathcal{M}$ and each child of a node represents a next step in the run of the ATM $\mathcal{M}$.

To mimic the start configuration, we define the following shapes:

$$h_1 \leftarrow Init \qquad s_{q_0} \leftarrow Init \qquad c_{d_i}^{(i)} \leftarrow Init \text{ for all } 1 \le i \le n$$

Intuitively, this is setting the starting state to $q_0$ (denoted by the shape name $s_{q_0}$), putting the head in the starting position ($h_1$), and stating the input word written on each tape cell ($c_{d_i}^{(i)}$).

The next step is to encode the transition relation of the $\mathcal{M}$. For each $1 \le i \le n$, each $(q, a) \in Q \times (\Sigma \cup \{B\})$, and $\gamma \in \{\ell, r\}$ we add the following shapes, where $(q', b, D) = \delta_\gamma(q, a)$:

$$s_{q'} \leftarrow \exists succ_\gamma^-.(s_q \wedge h_i \wedge c_a^{(i)}) \qquad c_b^{(i)} \leftarrow \exists succ_\gamma^-.(s_q \wedge h_i \wedge c_a^{(i)}) \qquad h_{i+D} \leftarrow \exists succ_\gamma^-.(s_q \wedge h_i \wedge c_a^{(i)}).$$

Furthermore, the tape cells that are not under the read-write head have their content preserved. Thus, for each $1 \le i \le n$, and $1 \le j \le n$ with $i \neq j$, add

$$c_a^{(i)} \leftarrow \exists succ^-.(c_a^{(i)} \wedge h_j).$$

We now identify subtrees that represent accepting computations. For all $q \in Q_\exists$ and all $q' \in Q_\forall$ we add the following:

$$s_{acc} \leftarrow s_{q_{acc}} \qquad s_{acc} \leftarrow s_q \wedge \exists succ.s_{acc} \qquad s_{acc} \leftarrow s_{q'} \wedge \exists succ_\ell.s_{acc} \wedge \exists succ_r.s_{acc}.$$

This concludes the reduction.

As we assumed $|\delta(q, a)| = 2$ for all $q \in Q, a \in \Sigma \cup \{B\}$, the run of $\mathcal{M}$ on an input word $w$ can be thought of as a binary tree with nodes in $\{0,1\}^*$ and edges $succ_\ell := \{(x, x0) \mid x \in \{0,1\}^*\}$ and $succ_r := \{(x, x1) \mid x \in \{0,1\}^*\}$, such that the root, $\epsilon$, represents the input word, combined with the starting position of the head and the name of the starting state; and each other node of this tree represents the next configurations of the run, such that if some node $x$ represents the word $e_1 \cdots e_n \in \Sigma^*$, with the head in position $i$ and state name $q$, such that $\delta_\ell(q, e_i) = (q', a', D')$, then the left-child of this node, $x0$, represents the word $e_1 \cdots e_{i-1} a' e_{i+1} \cdots e_n$, head position $i + D$ and state name $q'$, and similarly for the child on the right, $x1$, and $\delta_r(q, e_i)$.

Let $h$ be an homomorphism of $\{0,1\}^*$ into $can(\mathcal{T}, \mathcal{A})$. A simple check suffices to see that $x$ represents the word $e_1 \cdots e_n$, head position $i$ and state name $q$ iff $h(x)$ is decorated with the shape names $c_{e_i}^{(i)}$ for each $1 \le i \le n$, $h_i$ and $s_q$ in the perfect assignment.

An accepting run of $\mathcal{M}$ on the input word $w$ is a binary tree as described above such that there exists a (finite) subtree in which, if $x$ represents a state name $q \in Q_\exists \setminus \{q_{acc}\}$, then either $x0$ or $x1$ is part of the subtree, and if $x$ represents a state name $q \in Q_\forall \setminus \{q_{acc}\}$, then both $x0$ and $x1$ are part of the subtree. As we already concluded that $x$ represents a state $q$ iff $h(x)$ is decorated with $s_q$ in the perfect assignment, and a decoration with $s_{acc}$ requires a state to be an accepting state, or one successor to be decorated with $s_{acc}$ in case of $x$ representing a $Q_\exists$ state, resp. two in case of a $Q_\forall$ state, it is immediate that there exists such a subtree with all leaves representing the state name $q_{acc}$ iff we can decorate the root node $a$ in $can(\mathcal{T}, \mathcal{A})$ with the shape name $s_{acc}$. Thus, $\mathcal{M}$ accepts $w$ iff $can(\mathcal{T}, \mathcal{A})$ validates $(C, \{s_{acc}(a)\})$. ♣

## 10. Discussion and conclusion

*Beyond normalised Horn-$\mathcal{ALCHIQ}$.* In this article we considered Horn-$\mathcal{ALCHIQ}$ TBoxes in normal form. Unfortunately, we cannot easily lift the results to general Horn-$\mathcal{ALCHIQ}$, as our techniques are not immune to the usual normalisation procedures. For instance, the procedures in [23] and [35] may significantly change the form of the universal core model; the fresh concepts introduced during normalisation may force us to use different objects to satisfy axioms that would otherwise be satisfied by the same object. Take for instance the axiom $A \sqsubseteq \geq_2 r.B$ and the ABox $\mathcal{A} = \{A(a), r(a,b), B(b), r(a,c), B(c)\}$. In this case, $\mathcal{A}$ itself is a core universal model already. However, if we follow the normalisation as proposed in Lemma 2 of [23], we would find the following set of axioms, $\mathcal{T} = \{A \sqsubseteq \exists r.B_1, A \sqsubseteq \exists r.B_2, B_1 \sqsubseteq B, B_2 \sqsubseteq B, B_1 \sqcap B_2 \sqsubseteq \bot\}$. It is easy to check that in the core universal model $can(\mathcal{T}, \mathcal{A})$, $a$ has four outgoing $r$-edges to a $B$, instead of two. We believe it is possible to obtain similar results for full Horn-$\mathcal{ALCHIQ}$, albeit with a more cautious normalisation and a weaker notion of homomorphism. We leave the details for further work.

It would also be desirable to add transitivity axioms and thus cover the well-known Horn-$\mathcal{SHIQ}$ description logic. But unfortunately, in the presence of transitive roles, the uniqueness of the core universal model can no longer be guaranteed: take for instance $\mathcal{A} = \{A(a)\}$ and $\mathcal{T} = \{A \sqsubseteq \exists r.B, A \sqsubseteq \exists r.B', B \sqsubseteq \exists r.B', B' \sqsubseteq \exists r.B\}$ where $r$ is transitive. In this case, there are two (not strong) core universal models: a chain of $r$'s such that the concept names along this chain are, in the first node $A$, and then $B$ and $B'$, in an alternating fashion, closed under transitive roles, and the version in which $B$ and $B'$ swap places. Note that there exist (non-surjective) homomorphisms between both models. It is unclear what would be a good semantics if multiple core universal models exist.

*Towards full SHACL.* Extending the considered SHACL fragment is also a promising avenue. However, we first point that the 'guard' $c$ we introduced for constraints of the forms $\mathsf{eq}(E, r)$ and $\mathsf{disj}(E, r)$, cannot easily be dropped. Without it, the normalisation as described in the proof of Proposition 5.8 does not work well. There is no straightforward way to distinguish from which starting point it was possible to reach a final state, as becomes clear in the following example.

**Example 10.1.** Let $\mathcal{A} = \{r(a, c), t(b, c)\}$. Over this graph, we wish to check validation for the constraint $s \leftarrow \mathsf{eq}(r, t)$ with targets $s(a)$ and $s(c)$. Clearly, we expect non-validation. However, this cannot easily be achieved by a similar rewriting technique.

To see this, let us consider the following constraints as, loosely following the construction above to rewrite $s \leftarrow \mathsf{eq}(r, t)$ and based on the automata $\mathcal{M} = (\{q, q'\}, \{r, t\}, q, \{(q, r, q')\}, q')$ and $\mathcal{M}' = (\{q'', q'''\}, \{r, t\}, q'', \{(q'', t, q''')\}, q''')$ we find the following constraints:

$$s_q \leftarrow a \quad s_q \leftarrow c \quad s_{q'} \leftarrow \exists r^-.s_q \quad s_{pos} \leftarrow s_{q'} \quad s_{neg} \leftarrow \neg s_{q'} \quad s_{q''} \leftarrow a \quad s_{q''} \leftarrow c \quad s_{q'''} \leftarrow \exists t^-.s_{q''}$$
$$s_{pos} \leftarrow s_{q'''} \quad s_{neg} \leftarrow \neg s_{q'} \quad s_e \leftarrow s_{pos} \wedge s_{neg} \quad s_e \leftarrow \exists r.s_e \quad s_e \leftarrow \exists t.s_e \quad s_{noe} \leftarrow \neg s_e \quad s \leftarrow s_q \wedge s_{noe}$$

where we shortened the names of the error-shapes to $s_e$ and $s_{noe}$. As $b$ is reachable from starting states $a$ and $c$ by both an $r$ and a $t$, $b$ will only be assigned $s_{pos}$, the shape name denoting being the final state of at least one of the automata, not causing any error-shape $s_e$ to fire, which breaks our expectation of non-validation. The problem is that we cannot distinguish which initial state caused the positive outcome. ♡

For non-recursive SHACL, it is known that adding path equality or disjointness increases expressivity [24], but we do not know whether this also holds for recursive SHACL. In any case, it seems hard to include full path equality and disjointness due their 'non-local' nature.

On the other hand, it might be possible to encode local counting constraints of the form $\exists_{\leq n} r.\varphi$, after a careful examination of our techniques. Extending counting over roles to counting over regular paths is even more interesting. It will be quite a challenge to also define a rewriting in which the quadruples store information on counting over regular paths. However, as the anonymous parts of the core universal model are tree-like, it might be possible to also achieve such a rewriting.

In parallel to increasing the expressivity of the considered shape expressions, we may also consider richer targets. Additionally to shape atoms, the SHACL standard considers *target classes*: if a concept name $A$ is given as target, all nodes in the interpretation of $A$ must validate the given shape. For plain validation, without ontologies, class targets reduce trivially to sheer shape atoms: the set of domain elements is known and can replace the class or concept name. But if ontology axioms are given, this is a whole different story: the set of domain elements that forms the interpretation of the target concept name is not known a priori. Nevertheless, under the assumption that the input data graph is connected, it is possible to model this for $s \leftarrow \varphi$ with target class $A$, by creating new constraints of the form "there does not exists a path (the Kleene star of the union of all roles appearing) to an $A$ that does not satisfy $\varphi$". Now any domain element in the graph can be considered as the target to get a logically equivalent constraint. Clearly, a similar trick works if it is known which parts of the graph are connected.

Finally, we would like to emphasise that allowing full negation would be a very nice, but complicated challenge.

*Conclusion.* We have considered the validation of SHACL constraints in the presence of normalised Horn-$\mathcal{ALCHIQ}$ ontologies. To this end, we defined the semantics over a carefully constructed notion of a canonical model that minimises the number of fresh successors introduced to satisfy the ontology axioms at each chase step and which happens to be the unique core universal model. Moreover, we have argued that this semantics is natural and intuitive. We proposed a normal form and a rewriting algorithm for recursive SHACL constraints with *stratified* negation. It takes as an input a SHACL shapes graph and an ontology, and constructs a new SHACL shapes graph (also with stratified negation) that can be used for sound and complete validation over a slightly extended version of the data graph alone, without needing to reason about the ontology at validation time. We also discussed some approaches to even get to validation over the pure data graph. We showed that, under our semantics, validation in the presence of normalised Horn-$\mathcal{ALCHIQ}$ is complete for ExpTime, but it remains PTime complete in data complexity, and hence it is not harder than validation of stratified SHACL alone, without the ontology axioms.

## CRediT authorship contribution statement

**Anouk Oudshoorn:** Writing – original draft, Validation, Formal analysis, Conceptualization; **Magdalena Ortiz:** Writing – review & editing, Supervision, Methodology, Funding acquisition, Conceptualization; **Mantas Šimkus:** Writing – review & editing, Supervision, Methodology, Conceptualization.

## Data availability

No data was used for the research described in the article.

## Declaration of competing interest

The authors declare that they have no known competing financial interests or personal relationships that could have appeared to influence the work reported in this paper.

## Acknowledgements

The project leading to this application has received funding from the European Union's Horizon 2020 research and innovation programme under grant agreement No 101034440.

This work was partially supported by the Wallenberg AI, Autonomous Systems and Software Program (WASP) funded by the Knut and Alice Wallenberg Foundation. In addition, Šimkus was supported by the Austrian Science Fund (FWF) [10.55776/P30873], Ortiz was supported by the Austrian Science Fund (FWF) [10.55776/PIN8884924] and Oudshoorn was supported by the Netidee stipend 6794.